%% file: anonymous-submission-latex-2026.tex
\documentclass[letterpaper]{article} 
\usepackage{aaai2026}  
\usepackage{times}  
\usepackage{helvet}  
\usepackage{courier}  
\usepackage[hyphens]{url}  

\usepackage{graphicx} 
\usepackage{natbib}  
\usepackage{caption} 
\usepackage{algorithm}
\usepackage{algorithmic}
\usepackage{xcolor}
\usepackage{tikz}

\usetikzlibrary{positioning, arrows.meta}
\usepackage{newfloat}
\usepackage{listings}
\DeclareCaptionStyle{ruled}{labelfont=normalfont,labelsep=colon,strut=off} 
\floatstyle{ruled}
\newfloat{listing}{tb}{lst}{}
\floatname{listing}{Listing}
\title{Separating Clicks from Baits: Using Large Language Models to Detect Misleading YouTube Thumbnails}

\author{
    Wajiha Naveed\textsuperscript{\rm 1,2},
    Muhammad Muneeb Pervez\textsuperscript{\rm 2}\equalcontrib,
    Zaeem Mohtashim Khan\textsuperscript{\rm 2}\equalcontrib,
    Zafar Ayyub Qazi\textsuperscript{\rm 2,3},
    Zartash Afzal Uzmi\textsuperscript{\rm 2}
}

\affiliations{
    \textsuperscript{\rm 1}Duke University, Durham, North Carolina, USA\\
    \textsuperscript{\rm 2}Lahore University of Management Sciences, Lahore, Pakistan\\
    \textsuperscript{\rm 3}King Abdullah University of Science and Technology, Thuwal, Saudi Arabia\\
    wajiha.naveed@duke.edu,
    27100022@lums.edu.pk,
    26100107@lums.edu.pk,
    zafar.qazi@lums.edu.pk,
    zartash@lums.edu.pk
}

\usepackage{bibentry}

\newcommand{\answerTODO}{\emph}

\begin{document}

\maketitle

\begin{abstract}
Misleading video thumbnails on platforms like YouTube are a pervasive problem, undermining user trust and platform integrity. This paper proposes a novel multi-modal detection pipeline that uses Large Language Models (LLMs) to flag misleading thumbnails. We first construct a comprehensive dataset of 2,843 videos from eight countries, including 1,359 misleading thumbnail videos that collectively amassed over 7.6 billion views, providing a unique cross-cultural perspective on this global issue.
Our detection pipeline integrates video-to-text descriptions, thumbnail images, and subtitle transcripts to holistically analyze content and flag misleading thumbnails. Through extensive experimentation and prompt engineering, we evaluate the performance of four frontier-level LLMs, including GPT-4o, GPT-4o Mini, Claude 3.5 Sonnet, and Gemini-1.5 Flash. We further evaluate open-weight vision-language models, LLaVA-v1.5 and Qwen2.5-VL-7B-Instruct, to assess the generalizability of our approach beyond proprietary systems. Our findings show the effectiveness of LLMs in identifying misleading thumbnails, with Claude 3.5 Sonnet consistently showing strong performance, achieving an accuracy of 93.8\%, precision over 92\%, and recall exceeding 94\% in certain scenarios. Beyond evaluating detection performance, we conducted a careful failure analysis to understand when LLMs fail in identifying misleading thumbnails. We discuss the implications of our findings for content moderation, user experience, and the ethical considerations of deploying such systems at scale. Our findings pave the way for more transparent, trustworthy video platforms and stronger content integrity for audiences worldwide.
\end{abstract}

\begin{links}
    \link{Code and Dataset} {https://github.com/wajihanaveed/SCFBs.git}
\end{links}

\section{INTRODUCTION}
\input{introduction}

\section{METHODOLOGY}
\input{methodology}
\section{DATASET ANALYSIS}
\input{dataset-analysis}
\section{RESULTS}
\input{results}
\section{Comparison with Existing Work – CHECKER}
\input{checker}

\section{ABLATION STUDY}
\input{ablation}
\section{LIMITATIONS}

\input{limitations}
\section{DISCUSSION}
\input{discussion}

\section{RELATED WORK}
\input{related-work}

\section{CONCLUSION}
\input{conclusion}
\newpage
\bibliography{aaai2026}
\include{checklist}
\include{appendix}

\end{document}

%% file: introduction.tex
\label{sec:intro}
User-generated videos now dominate the web's information ecosystem. Platforms like YouTube, with over two billion monthly active users, serve as global hubs for information and entertainment. However, these platforms suffer from a persistent challenge: misleading thumbnail images. These deceptive previews are engineered to maximize clicks rather than accurately reflect content, undermining user trust and fueling clickbait and misinformation cultures.

Detecting misleading thumbnails is a critical problem with broad implications. Prior work~\cite{srinivasan2021impact} found that misleading thumbnails can boost click-through rates by 14\% compared to accurate ones, incentivizing their use. A Pew Research survey reported that 64\% of adults have encountered misleading content online, with thumbnails being a major contributor~\cite{pew2022misinformation}. Such practices not only erode user confidence but also distort content credibility and perceptions of reality. Establishing a responsible web ecosystem requires that addressing these vulnerabilities becomes a core priority for both systems engineering and platform governance.

However, addressing this issue at scale poses several challenges. Over 500 hours of video are uploaded to YouTube every minute~\cite{youtube2024press}, making manual review infeasible. Moreover, the definition of “misleading” is often subjective and culturally dependent, complicating automated detection. Traditional image recognition methods struggle to capture the semantic relationship between thumbnails and actual video content, leaving platforms reliant on limited automated tools and user reports, both of which fail to fully address the scope and nuance of the problem.

Against this backdrop, modern multimodal large language models offer a pragmatic solution. Such models’ capacity for joint reasoning over text, image, and video makes them well suited to assess semantic mismatches between a thumbnail and the underlying content. Prior work has shown that large language models exhibit strong zero-shot and few-shot generalization abilities across diverse tasks~\cite{brown2020languagemodelsfewshotlearners,wei2022emergentabilitieslargelanguage}, while multimodal vision-language models extend these capabilities to image and video understanding~\cite{tsimpoukelli2021multimodalfewshotlearningfrozen,liu2025fewshot}. Few-shot and retrieval-augmented prompting enable rapid adaptation across languages and cultural contexts without retraining, making these models a practical alternative for detecting misleading thumbnails at scale. Crucially, by evaluating open-weight vision-language models, we show that these capabilities also generalize beyond proprietary systems to accessible, non-closed platforms.

In this work, we evaluate whether such LLMs can reliably detect misleading thumbnails. We make the following key contributions:\\
\begin{enumerate}

\item \textbf{Comprehensive Dataset.} We compile a balanced dataset of 2,843 YouTube videos from eight countries, evenly split between misleading and non-misleading thumbnails. This enables support for cross-cultural evaluation of detection models across diverse content types. To ensure reproducibility and facilitate further study, we have made the dataset, annotation codebook, and scripts publicly available.

\item \textbf{Multi-Modal Analysis and Ablations.} We integrate video-to-text descriptions, thumbnails, and subtitles to holistically analyze content discrepancies, enabling a nuanced and accurate assessment of thumbnail deceptiveness. We run ablations to quantify the contributions of these input modalities.\\
\item \textbf{Multiple LLMs Evaluation.} We evaluate four frontier-level LLMs: GPT-4o, GPT-4o Mini, Claude 3.5 Sonnet, and Gemini-1.5 Flash, and complement this with two open-weight vision-language models: LLaVA-v1.5 and Qwen2.5-VL-7B-Instruct to assess performance across both proprietary and open ecosystems. \cite{llava15_13b,qwen2vl7b_instruct}\\
\item \textbf{Prompting Strategies} We study the effect of chain-of-thought reasoning, zero-shot, few-shot (fixed examples) and RAG-based dynamic few-shot techniques.\\
\item \textbf{Benchmarking against Task-Specific Solutions.} Our top configuration, Claude 3.5 Sonnet with dynamic few-shot prompts, was benchmarked against \textsc{Checker}~\cite{xie2021checker}, the leading supervised multimodal pipeline. Claude surpassed \textsc{Checker} on all metrics, showing that prompt-based LLMs can achieve state-of-the-art accuracy without task-specific training, offering a flexible, easily deployable alternative for content moderation.

\end{enumerate}

We evaluate model performance using four complementary metrics: accuracy, precision, recall, and specificity, to provide a balanced view of strengths. Our empirical results underscore the efficacy of LLM-based detection across diverse configurations. Claude 3.5 Sonnet emerges as the top-performing model, achieving a peak accuracy of 93.8\% under dynamic few-shot prompting while maintaining precision and specificity above 92\% and 93\%, respectively. While open-weight models such as Qwen2.5-VL-7B-Instruct demonstrate notable utility, reaching 74\% accuracy, proprietary closed-source systems, led by Claude 3.5 Sonnet, consistently define the performance ceiling. In several settings, both GPT-4o Mini and Claude 3.5 Sonnet exceed 94\% in recall, capturing most misleading thumbnails.

Our findings suggest that LLMs offer a promising path for detecting misleading thumbnails, with potential to enhance platform integrity and user trust. Real-world deployment will depend on tuning false-positive thresholds, ensuring transparent appeals, and adapting to evolving forms of manipulation. The paper details our methodology, ablation studies, model benchmarks, and implications for deployment, moderation, and user experience.


%% file: methodology.tex
This section describes the methodology used to study misleading thumbnails across multiple countries, including dataset construction, data processing, and analysis. We detail our process of country selection, YouTube video collection, thumbnail extraction, subtitle retrieval, video downloading, and descriptive text generation. These steps produce a comprehensive, reproducible dataset that supports large-scale analysis of misleading thumbnails and the use of LLM-based tools within the analysis pipeline.

\subsection{Country Selection}
To ensure a broad representation of content and cultural practices related to misleading thumbnails, we sampled videos from eight countries, four developing and four developed, drawn from the 20 nations with the largest YouTube audiences. Countries were classified by real GDP growth using the UN World Economic Situation and Prospects 2024 report~\cite{statista2024youtubeusers}. Brazil, Pakistan, Indonesia, and Mexico were categorized as developing countries, while the United States, the United Kingdom, Spain, and Italy were categorized as developed countries. 
This stratification enables a comparative analysis of misleading thumbnails across varying socio-economic contexts, specifically highlighting differences between high-income and middle-income digital landscapes.

\subsection{Dataset Construction}
We built a Misleading Thumbnail Videos (MTV) dataset using a multi-step process. VPNs simulated country-specific locations, and Google Chrome’s incognito mode minimized personalization. Guided by Google Trends, we used popular search terms per country. In Pakistan, trending searches revealed that single-character queries like “f” and “.” surfaced additional MTVs, so we included random-character searches to diversify results.
Videos were collected from both search results and the recommendation panel, leveraging YouTube’s tendency to recommend similar MTVs after viewing one \cite{hussein2020measuring}. For comparison, we also gathered Non-Misleading Thumbnail Videos (NMTVs). Non-English text in thumbnails or subtitles was translated with Google Translate.

The initial dataset contained 3,200 videos: 200 MTVs and 200 NMTVs from each of eight countries. Two annotators (trained graduates) labeled videos using a detailed codebook  (available in our GitHub repository) defining misleading thumbnails as those with exaggeration, false promises, or thematic mismatch. Minor exaggeration without thematic misrepresentation was labeled non-misleading. Annotators watched each video in full when the duration was under 5 minutes, and otherwise skimmed longer videos by focusing on the beginning, middle, and end to obtain a comprehensive overview of the content. Annotation time varied across videos but averaged approximately 6-7 minutes per video. Annotators reviewed both thumbnails and videos, achieving a Cohen’s Kappa $\kappa = 0.9633$ (near-perfect agreement). Only videos with full agreement were retained and items with disagreement were excluded from the final dataset. After removing disagreements, unavailable videos, and processing failures, the final dataset comprised 2,843 videos: 1,359 MTVs and 1,484 NMTVs.

\subsection{Data Processing}
To prepare our dataset for evaluation, we extracted three key modalities from each video, thumbnail image, subtitles, and a generated video-to-text description, capturing the cues a viewer encounters before or during early engagement.
We excluded social signals (comments, likes, views) to enable pre-hoc moderation, since these metrics are unavailable before upload and are often sparse or unreliable \cite{lindstol2023adolescents}. Video titles and descriptions were also omitted: prior work shows they rarely misrepresent content \cite{qu2018youtube}, and in our dataset deceptive titles were rare while descriptions were often empty or generic. In contrast, thumbnails, subtitles, and video-to-text summaries provide stronger, interpretable signals of thumbnail–content mismatch.

\vspace{0.5em}
\noindent \textbf{Thumbnail Extraction.} Thumbnails were downloaded via\\ \textit{https://img.youtube.com/vi/\{video\_id\}\/hqdefault.jpg}, stored on Google Cloud Platform for Gemini evaluations and locally for Claude, GPT-4o, and GPT-4o-mini.

\vspace{0.5em}
\noindent \textbf{Subtitle Retrieval.} We used a Python script with the YouTube Data API \cite{google2024youtubeapi} to fetch transcripts. Non-English subtitles were translated into English with Google Translate. Videos lacking auto-generated subtitles were kept for dataset consistency.

\vspace{0.5em}
\noindent \textbf{Video Download.} Videos were retrieved with the pytubefix library \cite{juanbindez2024pytubefix}. For videos over 30\,minutes, we analyzed only the first 29\,minutes and 55\,seconds to meet processing limits (e.g., Twelve Labs 30-minute cap, Gemini 1.5 Flash 50-minute cap). Videos were downloaded at 360p to save storage and uploaded to Google Cloud Platform, Twelve Labs, and local storage for processing.

\vspace{0.5em}
\noindent \textbf{Video Description Generation.}  
We generated video-to-text descriptions using Gemini \cite{google2024geminiflash}, Claude \cite{google2024claude35sonnet}, and Twelve Labs \cite{twelvelabs2024} to create structured summaries that capture key actions, visuals, and emotions, complementing subtitles for richer scene-level context. Instead of feeding raw video into classification prompts, we used concise descriptions to avoid accuracy loss from long inputs, which we observed on a smaller subset. This aligns with studies showing LLM performance degrades as context length grows \cite{databricks2024longrag,hsieh2024rulerwhatsrealcontext}. Descriptions balanced informativeness with token efficiency. \\
For Gemini and Twelve Labs, we prompted:
\begin{quote}
``Watch the video and provide a detailed description. Break down the content scene by scene, focusing on key actions, visuals, and emotions."
    
\end{quote}
This elicited temporally grounded narratives critical for interpreting thumbnails in context.
Claude 3.5 Sonnet, which lacks direct video input and accepts only 20 images, received 20 evenly spaced frames and the prompt:
\begin{quote}
   ``Consider these frames as continuous scenes from a video. Provide a detailed description of the video content, breaking it down scene by scene. Focus on key actions, visuals, emotions, and any notable details. Describe it as if you are watching the full video, ensuring that the narrative is cohesive and captures the flow of the scenes."
\end{quote}
Explicitly framing the stills as continuous scenes encouraged coherent, holistic narration comparable to video-aware models.
Video descriptions filled gaps when subtitles were sparse or absent and improved classification accuracy in ambiguous cases. For instance, a Ronaldo Photoshop tutorial with minimal subtitles was misclassified without the description, but correctly labeled once the description clarified context. 
Our ablation study confirmed that including descriptions consistently reduced misclassification and strengthened the multimodal pipeline.

\subsection{Prompts}

\begin{figure*}[t]
\centering
\includegraphics[width=0.85\textwidth]{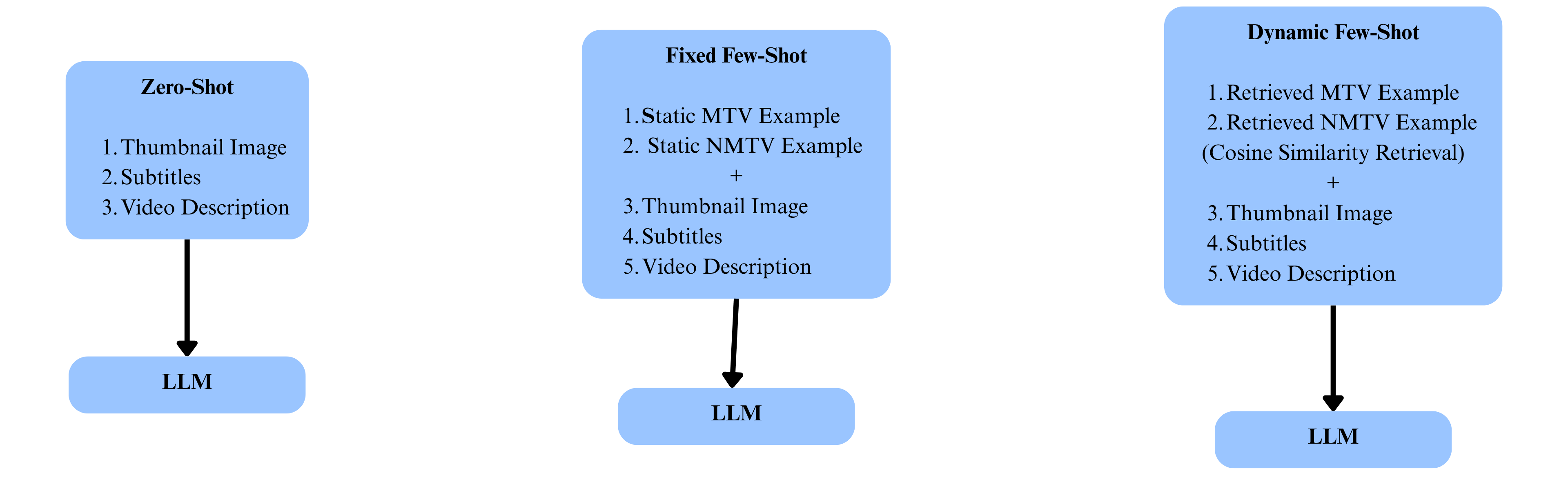}
\caption{Prompting strategies used for misleading thumbnail detection.}
\label{fig:prompting_strategies}
\end{figure*}

We explored various prompting strategies, as past research indicates that the structure and design of prompts significantly influence the reasoning performance of LLMs \cite{promptingguide2024techniques, kojima2023largelanguagemodelszeroshot}. Our experiments included three types of prompts: a Zero-shot prompt, followed by refinements into Fixed Few-shot and Dynamic Few-shot prompts. All three prompting strategies followed a clearly defined set of steps for classifying YouTube thumbnails. These instructions guided the LLMs in comparing the thumbnails with the actual video content and determining whether the thumbnails were misleading. Figure~\ref{fig:prompting_strategies} shows a contrasting difference between the three prompting strategies used.

\vspace{0.5em}
\noindent \textbf{Zero-Shot Prompt.}  
In the zero-shot setting, the LLM analyzed each video’s thumbnail, subtitles, and generated description, using a structured prompt with explicit step-by-step reasoning instructions. This follows the zero-shot chain-of-thought approach, where structured reasoning emerges without examples \cite{kojima2023largelanguagemodelszeroshot}.
To ensure consistent interpretation, the prompt defined a “misleading thumbnail” using YouTube’s policy, which describes thumbnails that “mislead viewers to think they’re about to view something that’s not in the video,” and prior research describing clickbait as content that “deviate[s] substantially from [the] content” it represents \cite{zannettou2018clickbait}. Our own annotation guidelines further refined this criterion.
For each video, the model compared the thumbnail to the subtitles and video description to decide whether the thumbnail accurately represented the main topic or used tactics such as exaggeration or false promises.
The full prompt template, which details these step-by-step instructions and input placeholders, is provided in the Appendix. This structured baseline prompt enabled a controlled evaluation of model performance before introducing few-shot or other advanced prompting strategies.

\vspace{0.5em}
\noindent \textbf{Fixed Few-Shot Prompt.}  
To improve consistency and reduce ambiguity, we extended the zero-shot prompt by adding two illustrative examples, one misleading and one not misleading, before the test instance. The prompt retained the same structure and step-by-step instructions but incorporated these examples to clarify how to distinguish thumbnails that exaggerate or misrepresent content from those that accurately reflect the video.
Each example, created collaboratively by the authors and LLMs to gauge the models’ own understanding, included a thumbnail description, a snippet of subtitles, a brief video description, the correct label, and a short explanation. These reference cases helped the model more reliably assess whether a test thumbnail aligned with the actual video content or employed clickbait tactics. The full example prompts are provided in the Appendix.

\begin{figure}[t]
\centering
\includegraphics[width=0.45\columnwidth]{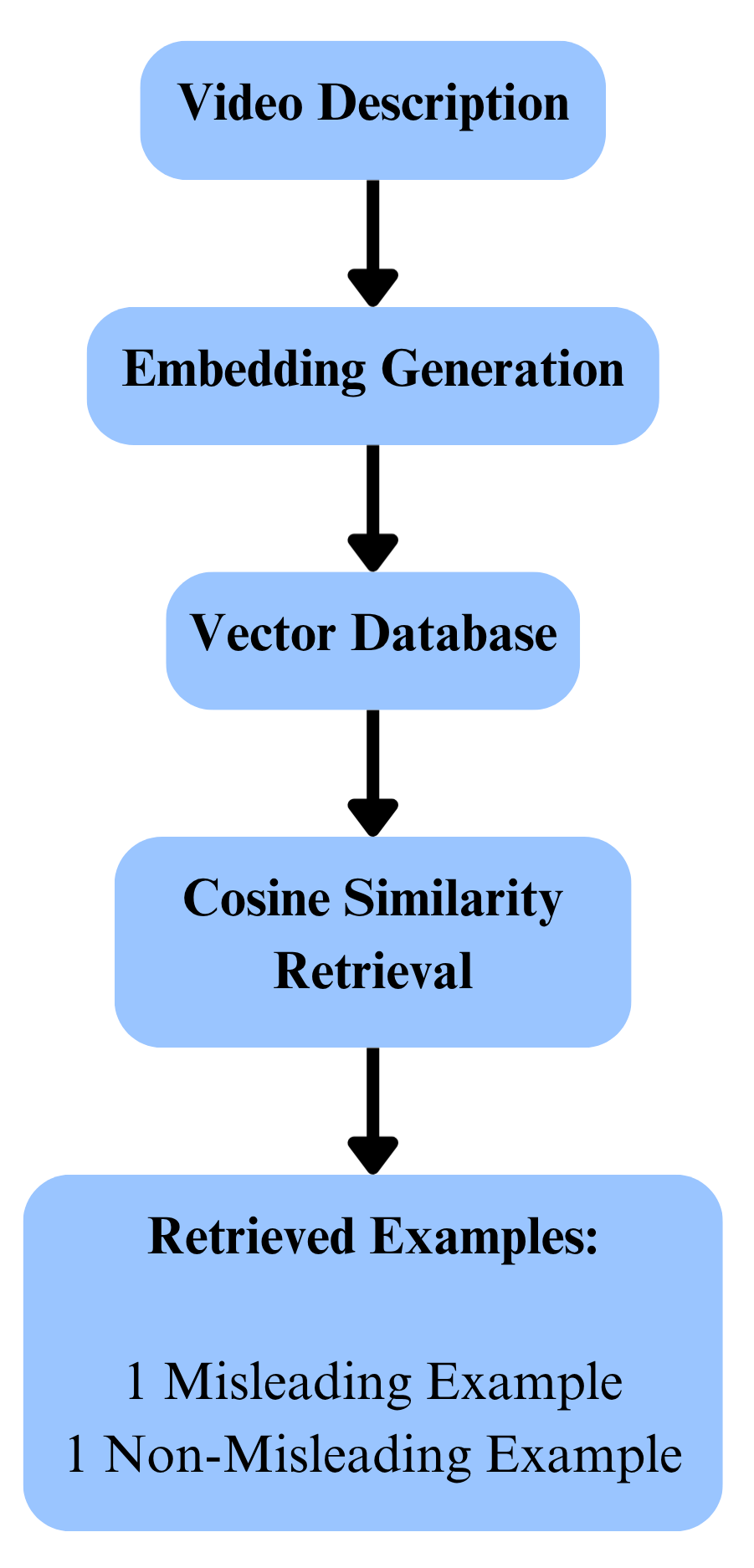}
\caption{Dynamic few-shot retrieval pipeline. Video descriptions are converted into embeddings and stored in a vector database. During inference, semantically similar examples are retrieved using cosine similarity.}
\label{fig:dynamic_retrieval}
\end{figure}

\vspace{0.5em}
\noindent \textbf{Dynamic Few-Shot Examples Prompt.} In the dynamic few-shot approach, we automatically selected two examples—one MTV and one NMTV—from the dataset that were semantically similar to the input video. As shown in Figure~\ref{fig:dynamic_retrieval}, we used a text-to-vector method, to analyze and compare video descriptions to ensure that the chosen examples closely related to the video under evaluation.\\

\begin{figure}[t]
\centering
{\hspace*{-0.6\columnwidth}
\includegraphics[width=2.2\columnwidth]{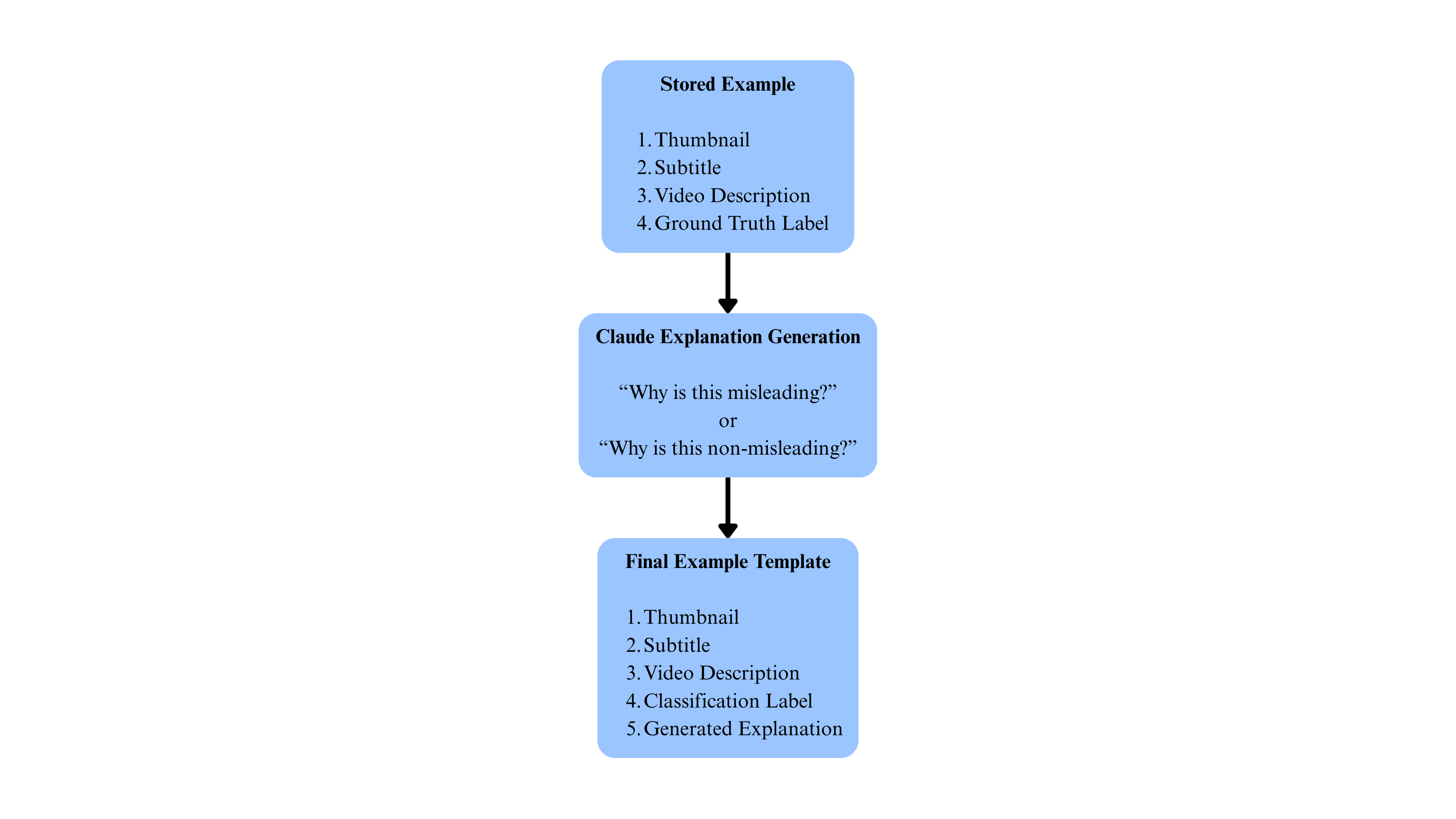}}
\caption{Construction of few-shot examples. Claude generates explanations describing why misleading thumbnails are misleading and vice versa.}
\label{fig:example_generation}
\end{figure}

\begin{enumerate}

\item 
\textbf{Text-to-Vector Conversion and Similarity Analysis.}
In the dynamic few-shot approach, we automatically selected two semantically similar examples, one MTV and one NMTV from the dataset to accompany each test instance. Video descriptions were embedded into vectors using Sentence-BERT (SBERT), which efficiently produces high-quality sentence embeddings via a Siamese architecture \cite{reimers2019sentencebertsentenceembeddingsusing}. Unlike standard BERT or RoBERTa, SBERT is highly efficient, reducing similarity search time among 10,000 sentences from ~65 hours to just 5 seconds while preserving semantic accuracy. This makes it well-suited for retrieval tasks like our dynamic few-shot prompting setup. Cosine similarity identified the most contextually similar videos, enabling balanced retrieval of misleading and non-misleading examples.
To ensure fairness, all video descriptions were uniformly generated with Twelve Labs, which was not used for classification. This avoided model-specific bias and provided consistent semantic cues.
\item \textbf{Thumbnail Descriptions and Explanations.}
We precomputed the thumbnail descriptions for all thumbnails in our dataset. Claude was selected to generate concise, one-sentence descriptions for each thumbnail due to its high accuracy in similar tasks. Since the thumbnails themselves were not included in the prompt, these descriptions served as the textual representation.
Next, we generated explanations for why a video’s thumbnail was categorized as misleading or not using Claude. With the thumbnail descriptions, ground truth labels, and truncated video descriptions and subtitles (both limited to 200 words), the model produced concise rationales for each classification. As shown in Figure~\ref{fig:example_generation}, these explanations were incorporated into a standardized example template used for dynamic few-shot prompting to ensure consistency in the evaluation process.
\end{enumerate}











\noindent \textbf{Prompt Integration.}
Each example followed a standardized format, featuring a textual description of the video’s thumbnail, truncated versions of both the video’s subtitles and video description (limited to 200 words each), and a categorization label specifying whether the thumbnail was ``Misleading" or ``Not Misleading," accompanied by an explanation. These examples helped the model evaluate and categorize thumbnails more accurately and consistently. The 200-word limit for subtitles and video descriptions was set to avoid overwhelming the model, as longer inputs significantly reduced the accuracy.

\begin{figure*}[t]
\centering
\includegraphics[width=\textwidth]{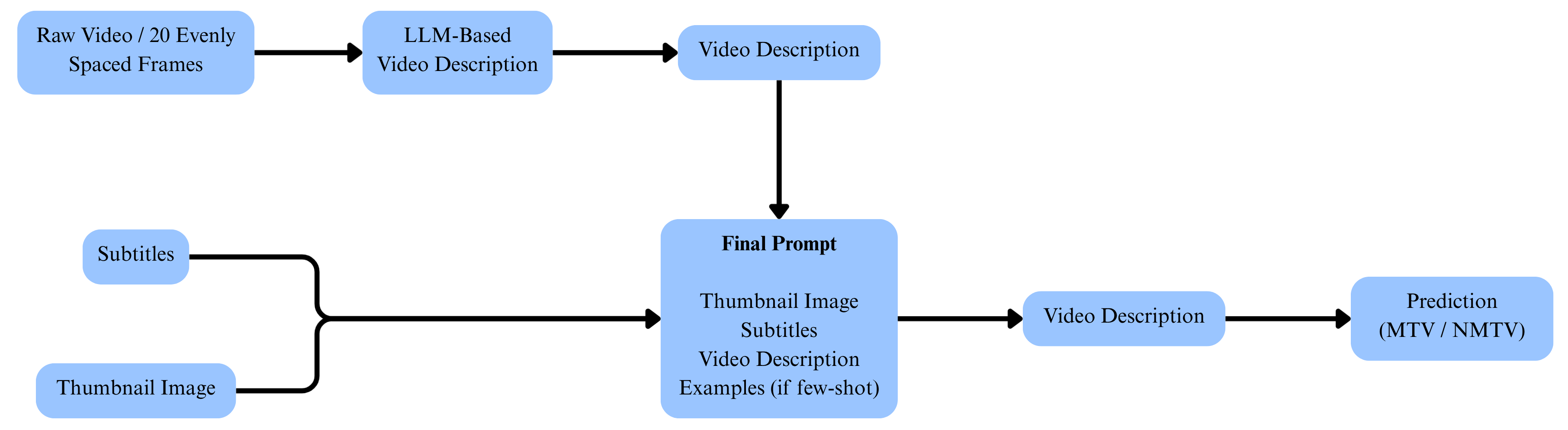}
\caption{Overall misleading thumbnail detection pipeline. The framework combines thumbnail images, subtitles, video descriptions, and prompting strategies to classify thumbnails as misleading (MTV) or non-misleading (NMTV).}
\label{fig:overall_pipeline}
\end{figure*}








\subsection{Models and Pipelines}


We evaluate multiple multimodal pipelines using identical inputs (subtitles, thumbnails and supporting metadata), differing only in how video descriptions are generated and the underlying model architecture. Figure~\ref{fig:overall_pipeline} illustrates the overall system architecture and evaluation pipeline used across all models.

Two of the four proprietary models we evaluated, namely \texttt{claude-3-5-sonnet@20240620} and \texttt{gemini-1.5-flash-001} use their own generated video descriptions while \texttt{gpt-4o-mini} and \texttt{gpt-4o} \cite{openai2024models} use video descriptions generated by Twelve Labs. For \texttt{gpt-4o}, we use the default release \texttt{gpt-4o-2024-05-13}. This is because, due to resource constraints, video descriptions for the GPT models were generated using Twelve Labs, which provided platform-supported credits for scalable video processing. These models were executed with default temperature settings and a maximum token limit of 4{,}800 tokens.

In addition, we evaluate open-weight vision–language models, LLaVA-1.5 (13B) and Qwen-2.5-VL  (7B), using the same prompting pipeline and multimodal inputs. Due to resource constraints, video descriptions generated using \texttt{claude-3-5-sonnet@20240620} were used to ensure high-quality inputs.

This unified setup enables a controlled comparison across models by holding inputs and prompting strategies constant while varying model capabilities. It allows us to evaluate trade-offs between performance, cost and accessibility, highlighting differences between closed-source systems with native multimodal integration and open-weight alternatives that offer greater transparency and reproducibility.

%% file: dataset-analysis.tex
This section presents an analysis of our curated dataset of 2,843 videos, comprising 1,359 MTVs and 1,484 NMTVs from eight countries. We examine category distributions, analyze the prevalence of various misleading tactics, and critically reflect on the dataset’s representativeness, including the measures taken to mitigate potential collection biases.

\subsection{Video Categories and Distribution}

\begin{figure}
\centering

\includegraphics[width=\linewidth]{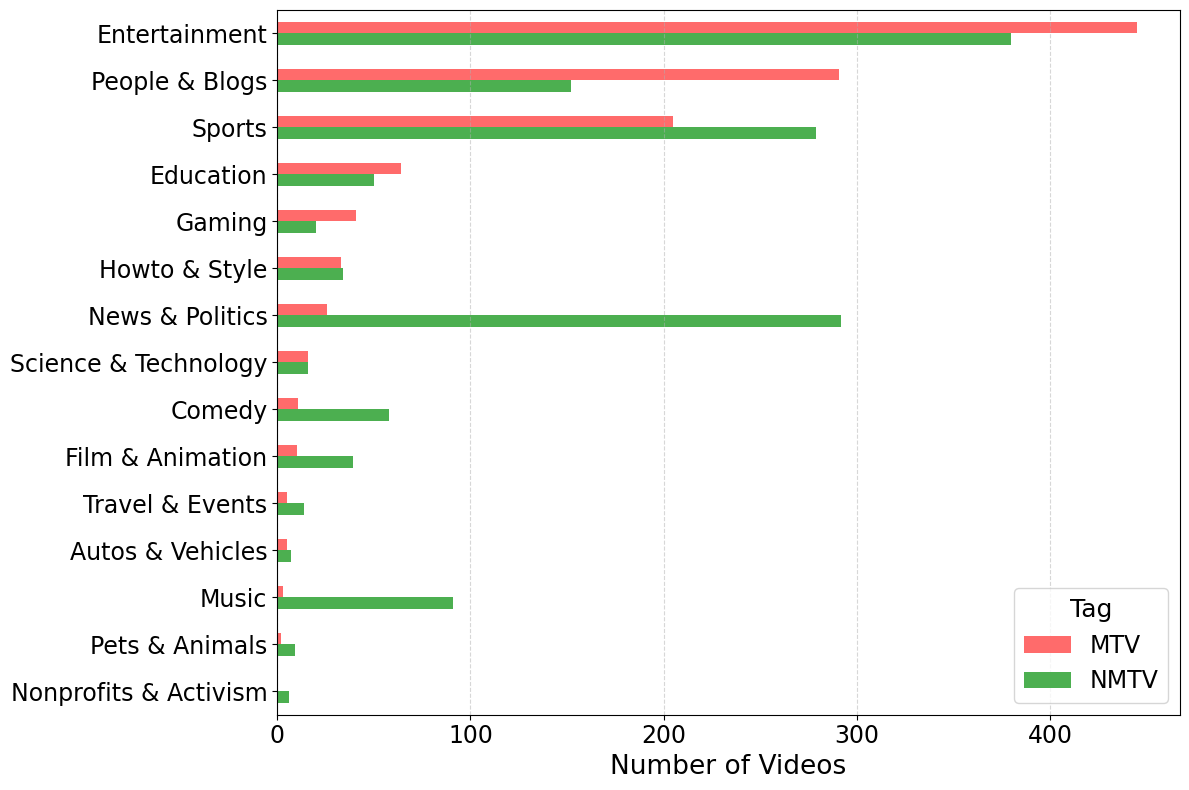}
\caption{Overall categorical distribution of MTVs and NMTVs.}

\label{fig:all_cat}
\end{figure}

Figure~\ref{fig:all_cat} shows the distribution of video categories in our dataset. The top three, \textit{Entertainment}, \textit{Sports}, and \textit{People \& Blogs}, dominate both MTVs and NMTVs, aligning with broader YouTube trends where these high-engagement categories are frequently linked to misleading thumbnails. Their prominence highlights the need for detection strategies focused on high-volume content.
Despite this global dominance, regional differences emerge. In developed regions, \textit{Entertainment} leads for both MTVs and NMTVs, followed by \textit{People \& Blogs} and \textit{Sports}, with other categories trailing off, reflecting an emphasis on personal, lifestyle, and culturally resonant media. In contrast, developing countries show a more diverse distribution. Although the top three categories remain prevalent, \textit{News \& Politics} is notably more prominent among NMTVs, which might be due to stricter moderation or editorial practices \cite{doi:10.1177/14614448231213942}. Categories like \textit{Music}, \textit{Pets \& Animals}, and \textit{Science \& Technology} also appear with balanced, moderate frequency, indicating varied content interests.
These patterns emphasize the need for region- and category-aware detection strategies that incorporate not only visual but also cultural and contextual cues. Our pipeline supports both region-wise and category-wise analysis, enabling more robust and generalizable detection of misleading thumbnails across diverse global content ecosystems.

\subsection{Dataset Bias and Selection Strategy}

We acknowledge that certain video types, particularly MTVs from entertainment-driven channels, are overrepresented in our dataset. However, this skew is not arbitrary; it reflects the real-world prevalence of misleading thumbnail tactics in high-traffic genres such as entertainment and sports. Our aim was not to replicate YouTube’s global content distribution, but rather to capture misleading behavior where it naturally occurs at scale.

Since MTVs were identified through manual or “accidental” discovery methods, such as trending content or random keyword searches, this process reflects how misleading content typically surfaces on the platform. Rather than artificially flattening category distributions, we preserved these natural patterns, which align with documented engagement trends on YouTube.

Prior work has often relied on datasets heavily skewed toward NMTVs, limiting the evaluation of false negatives. Although our dataset does not fully represent YouTube’s overall video ecosystem, its balanced design enables rigorous testing across both MTVs and NMTVs, providing a reliable benchmark for evaluating misleading thumbnail detection.




\subsection{Approaches to Misleading Thumbnail Design}
Our analysis revealed various tactics used to create misleading thumbnails on YouTube. These tactics can be grouped into distinct categories.\\
\begin{enumerate}
\item \textbf {Exaggeration Tactics.} Thumbnails often exaggerate "before and after" scenarios, such as rapid weight loss or anti-aging.\\
\item \textbf{Celebrity Manipulation.} Celebrities are often shown in compromising settings, like jail or hospitals, with fake dialogue bubbles conveying strong emotions.\\
\item \textbf{Lifestyle Fantasies.} Many thumbnails showcase exaggerated luxurious lifestyles like cars, mansions and private planes, misleading viewers into believing the video content will mirror those images.\\
\item \textbf{Fabricated Visuals.} Some thumbnails use manipulated images, such as merging human and animal features, or bold claims like ``married" or ``divorce confirmed" that are not substantiated by the video content.\\
\item \textbf{Provocative and Sensational Language:} Words like ``exclusive," curse words, and similar attention-grabbing terms are used, often without proper context.\\
\item \textbf{Regional Trends.} In certain regions, a unique trend has emerged where users search for videos using only a period (full stop). These ``full stop" videos often feature disturbing or creepy thumbnails, part of a meme-like search behavior on YouTube.
\end{enumerate}

\subsection{Effectiveness of YouTube's Thumbnail Policy}
YouTube has a policy in place for handling misleading thumbnails, which can lead to their removal or, in more severe cases, the termination of an entire channel \cite{youtube2024thumbnails}. YouTube relies on user reports to flag these thumbnails, in addition to employing machine learning algorithms for detection \cite{youtube2024communityguidelines}. However, many misleading thumbnails go unreported by users, limiting the effectiveness of the current system. From our dataset of 1,359 MTV videos, the average video age was 442 days. Of these, the top 10 most viewed videos had an average age of 924 days, and only 65 videos were removed from the entire dataset over the course of seven weeks, highlighting the inefficiency of this approach in addressing the issue at scale.

We evaluated how LLMs detect misleading YouTube thumbnails by comparing performance across prompt types, measuring the number of videos processed, and analyzing prediction accuracy. We also performed a cross-country comparison to examine regional variations in detection performance and benchmarked our results against existing standards. Finally, we assessed the computational costs of each model to identify the most accurate and cost-efficient solution for large-scale thumbnail analysis.

%% file: results.tex
\subsection{Variation in Number of Processed Videos}

The number of processed misleading thumbnail videos varied across models due to differences in filtering. Google’s Gemini 1.5 Flash applied strict safety filters, excluding content flagged under enum codes like \texttt{PROHIBITED\_CONTENT} (e.g., sensitive topics), \texttt{SAFETY} (e.g., hate speech), and \texttt{RECITATION} (unauthorized citations)~\cite{google2024geminisafety,google2024safetyfilters}.

In contrast, Claude 3.5 Sonnet, GPT-4o, and GPT-4o Mini employed more permissive safety filters~\cite{openai2024safety,anthropic2024aup}, allowing a broader but riskier video set. Twelve Labs lacked explicit safety filters but excluded videos under 360p resolution, limiting coverage in regions with low-quality uploads.

These disparities introduce a study limitation: model-level variation in filtering and resolution constraints directly affected video coverage. More concerningly, the accessibility of policy-violating videos, despite YouTube’s moderation framework suggests upstream issues. Ideally, such content should be blocked at upload, not filtered post-hoc via third-party tools. Improving safety mechanisms at both platform and model levels remains an open challenge~\cite{kumar2024ethics,li2024safety}. A full breakdown of processed videos per model appears in Table~\ref{tab:videos_processed} (Appendix).
\subsection{Comparison Across Proprietary Models}

\begin{figure}
\centering
\vspace{-0.3in}
\includegraphics[width=\linewidth]{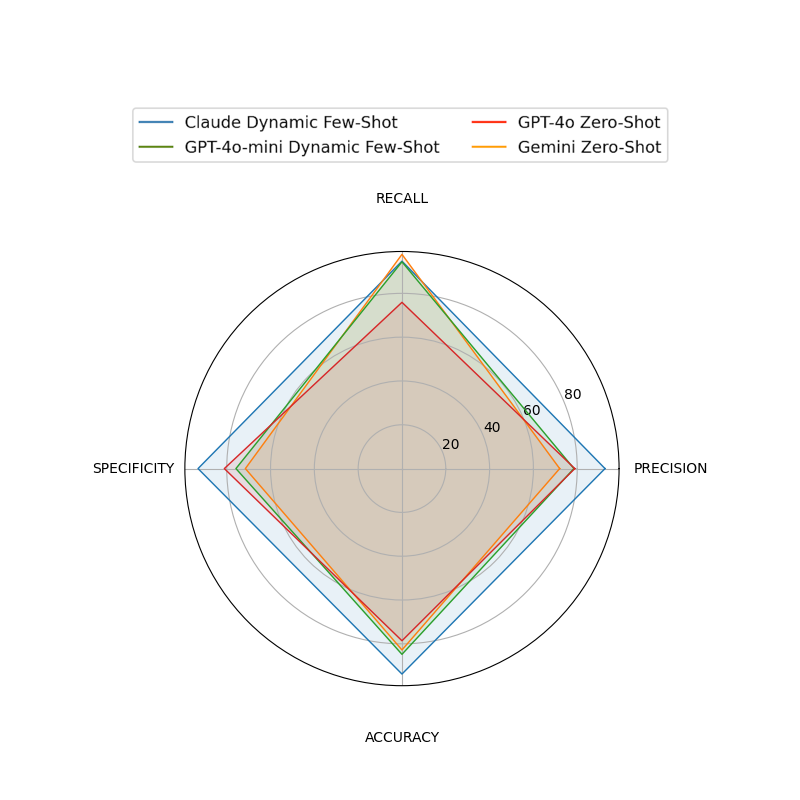}
\vspace{-0.4in}
\caption{Radar plot of best prompt accuracy for four models across Accuracy, Recall, Precision, and Specificity.}
\label{fig:radar_all}
\end{figure}

\noindent We evaluated the four models, Claude 3.5 Sonnet, Gemini 1.5 Flash, GPT-4o-mini, and GPT-4o, across accuracy, precision, recall, and specificity. Figure~\ref{fig:radar_all} and Table~\ref{tab:model_accuracy} highlight key differences in their ability to detect misleading thumbnails.

\begin{table}[h!]
\centering
\begin{tabular}{lc}
\hline
\textbf{Model} & \textbf{Accuracy (\%)} \\ \hline
Claude 3.5 Sonnet & 93.8 \\
Gemini 1.5 Flash & 82.8 \\
GPT-4o Mini & 84.8 \\
GPT-4o & 78.6 \\
\hline
\end{tabular}
\caption{Best accuracy of models across prompting strategies.}
\label{tab:model_accuracy}
\end{table}
\noindent \textbf{Claude 3.5 Sonnet} achieved the highest accuracy (93.8\%), driven by its strong chain-of-thought (CoT) reasoning even without explicit CoT prompting. It consistently generated structured outputs that detected visual-textual mismatches, particularly in borderline cases such as distinguishing interviews from dramatizations. A representative example is provided in Appendix~\ref{appendix:claude_response}. While Claude outperformed other models overall, all models, including Claude, struggled with thumbnails that promised full movies or episodes (e.g., “Watch Complete Movie” or “Full Episode”) but actually contained only image slideshows or unrelated voiceovers. In these cases, they failed to detect the mismatch between the thumbnail’s claim and the video’s actual content.

\noindent \textbf{Gemini 1.5 Flash} achieved an accuracy of 82.8\% and excelled in recall (97.8\%) but lagged in specificity (71.5\%), frequently over-flagging non-misleading content. It exhibited weak contextual understanding, particularly due to poor celebrity recognition. For example, it misclassified a thumbnail featuring Angelina Jolie because it failed to align the visual content with the descriptions.\\
\noindent \textbf{GPT-4o-mini} achieved an accuracy of 84.8\%. While it trailed Claude, it outperformed Gemini on complex videos, showing a stronger grasp of subtle visual cues. Some errors stemmed from failing to detect edited romantic thumbnails. Although it misclassified fewer cases than Gemini, it still struggled to distinguish real from fabricated thumbnails.

\noindent \textbf{GPT-4o} performed worst overall (78.6\% accuracy). It frequently missed key contextual or visual cues such as failing to identify Lionel Messi in a fabricated thumbnail and often defaulted to surface-level interpretations. It also misclassified exaggerated or surreal thumbnails (e.g., a carved watermelon eagle) as legitimate.

\vspace{0.5em}
\noindent \textbf{Trade-offs in Metrics.} Beyond accuracy, precision and specificity revealed critical deployment trade-offs. Claude achieved the highest specificity (0.931), minimizing false positives—a key factor in user trust. In contrast, Gemini's high recall came at the cost of precision, flagging benign content more often. GPT-4o-mini offered moderate but balanced performance, while GPT-4o underperformed across all metrics.

\vspace{0.5em}
\noindent \textbf{Key Differentiators.} Claude stood out for its ability to reason through emotionally charged or visually exaggerated content. Gemini was strong on recall but weak in context comprehension. GPT-4o-mini handled nuanced prompts better than GPT-4o, which frequently missed critical details.

\vspace{0.5em}
\noindent \textbf{Limitations.} Across models a few weaknesses stood out: difficulty detecting misleading “full movie” or “full episode” thumbnails and poor celebrity recognition, which weakened contextual understanding. Claude handled most of these cases better through inference but still misclassified some. Addressing these gaps will require stronger vision–language grounding and entity recognition.

\vspace{0.5em}
\noindent \textbf{Summary.} Claude sets the benchmark in accuracy, reasoning, and reliability. GPT-4o-mini shows potential in nuanced classification, Gemini excels in recall but lacks precision, and GPT-4o remains limited to simpler tasks. Improving visual reasoning and celebrity-aware understanding will be essential for next-generation models.

\subsection{Comparison Across Prompts}

\begin{figure}
    \centering
    \includegraphics[width=\linewidth]{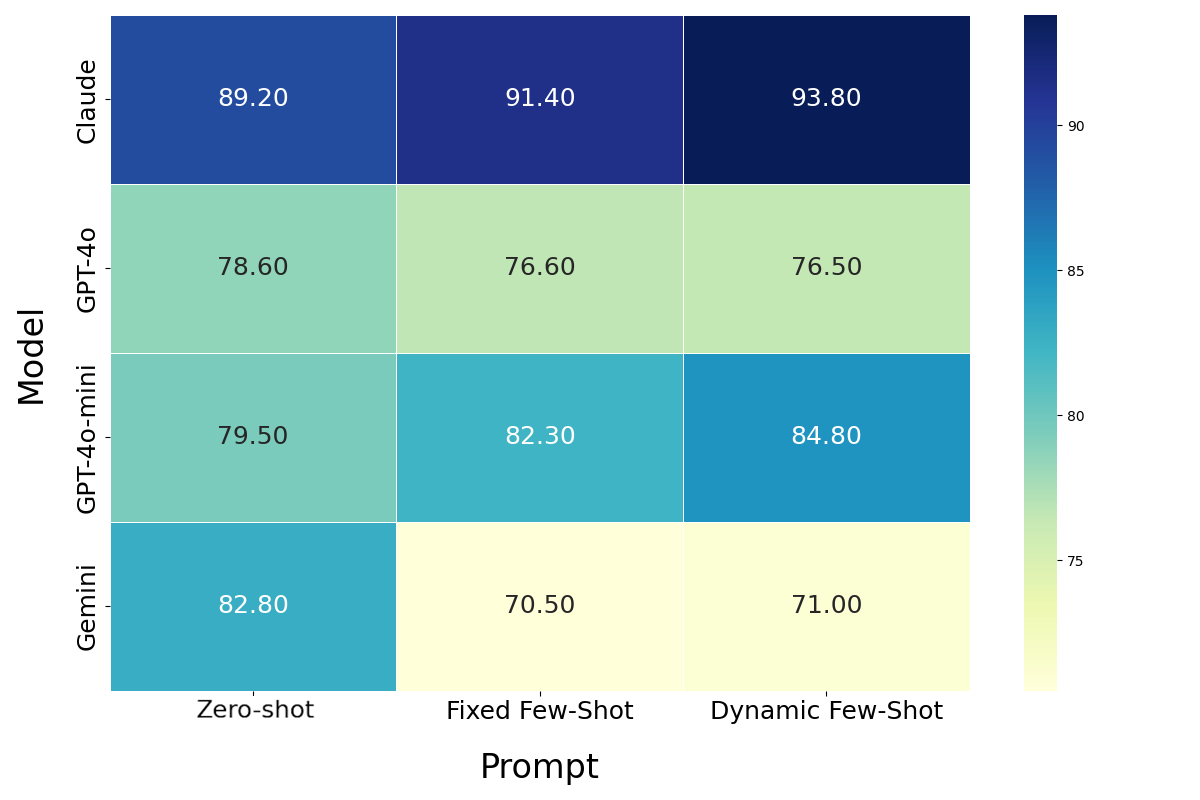}
    \caption{Model accuracies for each prompt}
    \label{fig:heat_map}
\end{figure}

We observed notable differences in accuracy when using different prompting strategies within the same model. Our hypothesis was that transitioning from \textit{zero-shot} to \textit{fixed few-shot} and, finally, to \textit{dynamic few-shot} would lead to increased accuracy and improved overall metrics, as prior research suggested. This trend was evident in the top two models, Claude 3.5 Sonnet and GPT-4o-mini, which displayed consistently higher accuracies across prompts as observed in Figure \ref{fig:heat_map}.

For Claude 3.5 Sonnet, fluctuations in accuracy were observed with \textit{zero-shot} prompting, with some regions falling below 85\% (see Appendix Figure \ref{fig1}). However, accuracy improved significantly with the introduction of \textit{few-shot} and \textit{dynamic prompting}. These improvements demonstrate the value of advanced prompting strategies, particularly when working with complex input settings. By using these refined techniques, Claude's accuracy consistently exceeded 90\% across all regions.

In contrast, GPT-4o's performance remained relatively static across different prompting strategies, showing little improvement when moving from \textit{zero-shot} to \textit{few-shot} prompting. Meanwhile, Gemini 1.5 Flash displayed a notable decrease in accuracy when moving from \textit{zero-shot} to \textit{few-shot} prompts, as shown in Figure \ref{fig:heat_map}. This result aligned with past research and our preliminary tests, which indicated that Gemini, while generally less effective than other models, performed relatively better when given simpler instructions, such as determining whether a video was misleading or not \cite{ataallah2024infinibench}. Although the performance declined with more detailed prompts, these findings provided valuable insights into the model’s behavior and highlighted areas for improvement in future applications.

\subsection{Performance Across Categories}

To evaluate how well the detection pipeline generalizes across diverse video categories, we analyzed the performance of Claude 3.5 Sonnet (with dynamic prompting) on a balanced subset of categories. We addressed class imbalance by sampling an equal number of MTVs and NMTVs for each category, using a 1:1 ratio based on the smaller class size (i.e., \texttt{min(total\_MTV, total\_NMTV)}), thereby ensuring fairness while preserving category diversity. Categories with no MTVs, such as \textit{Pets \& Animals} and \textit{Nonprofits \& Activism}, and those with very limited data ($\le 10$ videos after balancing, e.g., \textit{Autos \& Vehicles}) were excluded to maintain metric reliability and avoid misleading conclusions. Although most of these categories showed promising results, their low support made the metrics unreliable.

\begin{table}[h]
\centering
\begin{tabular}{lcc}
\hline
\textbf{Category Name} & \textbf{Accuracy} & \textbf{F1 Score} \\
\hline
Sports & 0.9530 & 0.9535 \\
Gaming & 0.9474 & 0.9500 \\
Education & 0.9388 & 0.9412 \\
Entertainment & 0.9107 & 0.9108 \\
Comedy & 0.9091 & 0.9091 \\
Howto \& Style & 0.9062 & 0.9032 \\
News \& Politics & 0.9038 & 0.8936 \\
Film \& Animation & 0.9000 & 0.8889 \\
People \& Blogs & 0.8826 & 0.8875 \\
Science \& Technology & 0.8750 & 0.8667 \\
\hline
\end{tabular}
\caption{Per-Category Accuracy and F1 Scores (Balanced Dataset)}
\label{tab:num_category_scores}
\end{table}

As shown in Table~\ref{tab:num_category_scores}, the model demonstrates consistently strong performance across categories, with accuracy ranging from 0.8750 to 0.9530 (a spread of 7.8\%). The highest-performing category, \textit{Sports} (95.3\%), outperforms the lowest-performing category, \textit{Science \& Technology} (87.5\%), by 7.8 percentage points. Similarly, F1 scores range from 0.8667 to 0.9535 (an 8.7\% spread), indicating stable performance across domains.

High-performing categories such as \textit{Sports} and \textit{Gaming} exceed 94\% accuracy, while mid-tier categories like \textit{Entertainment} and \textit{Comedy} remain around 91\%, representing a modest drop of approximately 3--4 percentage points. Lower-performing categories, including \textit{People \& Blogs} and \textit{Science \& Technology}, fall below 88\%, suggesting increased ambiguity or weaker visual cues for misleading content.

Overall, the relatively narrow performance gap ($<10\%$ across all categories) highlights the pipeline’s robustness, while still indicating room for improvement in more nuanced or less visually explicit domains.

\subsection{Analysis of Misleading Thumbnails Across Countries}

\begin{figure} 
\centering 
\includegraphics[width=\linewidth]{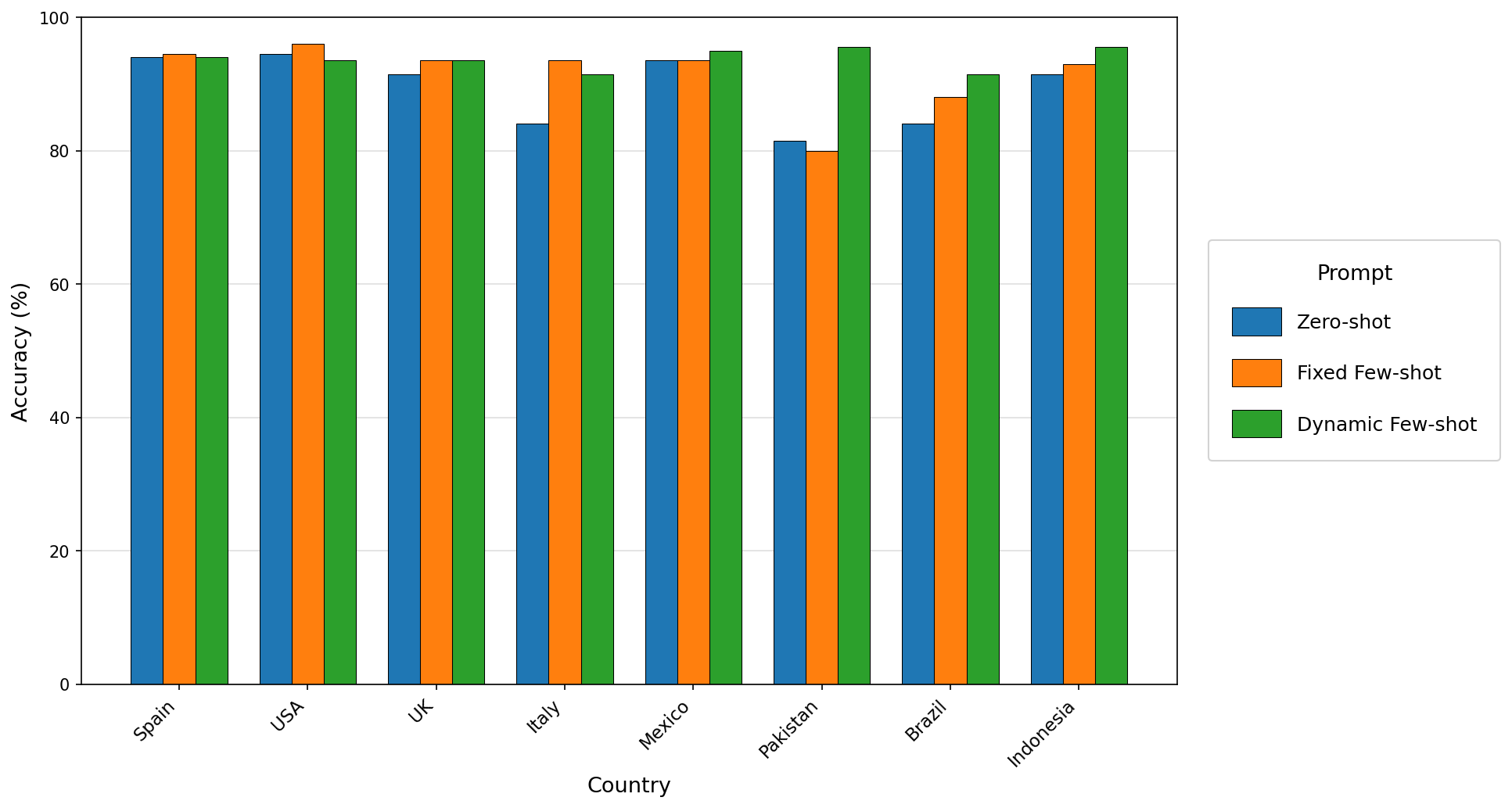} 
\vspace{-0.3in}
\caption{Accuracy using Claude 3.5 Sonnet for each prompt across all countries.
} 
\label{fig:bestaccuracy_overall} 
\end{figure}

Model performance varied across countries. On average, MTV detection accuracy was similar between developed (82.3\%) and developing (80.6\%) regions. However, Claude's zero-shot prompt showed substantial country-level variation: some countries exceeded 93\% accuracy, while others such as Italy, Brazil and Pakistan remained closer to 80\%, resulting in a performance gap of over 13 percentage points between the best and worst-performing regions.
Misclassifications often stemmed from thumbnails with exaggerated or sensational localized content. In Pakistan, a distinct subset of MTVs featured local celebrities in dramatized scenarios, particularly from the film and cricket industries—highlighting localized clickbait styles. These findings suggest that cultural, linguistic, and design differences in thumbnails affect LLM generalization. To address this, we used dynamic few-shot prompting with culturally relevant examples. As shown in Figure~\ref{fig:bestaccuracy_overall}, this led to substantial gains; accuracy improved by at least 8 percentage points in previously low-performing countries, reducing the cross-country performance gap from over 13\% to under 3\%, with all countries surpassing 90\% accuracy.
This underscores the importance of culturally adapted inputs for robust cross-regional performance.

\vspace{0.5em}
\noindent \textbf{Common Strategies.} Despite regional variation, several tactics were widespread across all countries. Celebrities were heavily used in misleading thumbnails, with Pakistan and Spain primarily featuring athletes, while the United States emphasized YouTubers and film actors. Thumbnails frequently exaggerated wealth or personal transformation, appealing to aspirational themes. In addition, provocative or visually unrelated imagery was commonly employed to attract user attention and increase click-through rates.

\vspace{0.5em}
\noindent\textbf{Key Differences.} Notable cross-regional differences emerged in how these strategies were deployed. In terms of sensationalism, developing countries more often relied on fantastical or implausible scenarios, whereas developed countries favored subtler forms of exaggeration, such as insinuations of celebrity drama. Sexualized content was more overt in developing regions, often pushing against prevailing societal norms, while in developed regions it appeared in more satirical or dramatized forms. Finally, portrayals of wealth and status varied substantially: exaggerated displays of wealth were particularly prominent in countries such as Pakistan, whereas emotional turmoil and celebrity disputes were more commonly emphasized in the United States. In summary, detecting misleading thumbnails at scale requires recognizing regional strategies and cultural cues. Incorporating localized data and prompt adaptation significantly boosts performance across diverse global contexts.
\subsection{Benchmarking and Performance Discrepancies in Proprietary LLMs}
Our results align with existing works on model benchmarks, confirming Claude 3.5 Sonnet's superior performance in classification tasks. Claude consistently demonstrated the highest accuracy in our study, with an average of 91.5\% across multiple prompts, maintaining low error rates even when handling complex and misleading thumbnails \cite{arshad2024ageval,Whitbeck_2024}. 

Gemini 1.5 Flash, while highly effective in blocking harmful content, processed fewer videos due to its strict safety filters, resulting in an average accuracy of 74.8\%. GPT-4o and GPT-4o-mini followed with average accuracies of 77.3\% and 82.2\%, respectively, showing competence but needing further improvement in managing complex or provocative thumbnails. Notably, GPT-4o-mini outperformed GPT-4o by 4.9 percentage points, a counter-intuitive result given GPT-4o’s larger capacity. Prior studies similarly report GPT-4o-mini’s stronger performance over GPT-4o and Gemini in intent classification and knowledge-based reasoning tasks \cite{maheshwari2024efficacy,sinha2024guiding}, suggesting that smaller or more task-aligned models may generalize better in structured classification settings.

Although Gemini excels in localized tasks such as temporal reasoning and summarization, it struggles with more complex, global tasks requiring deeper context understanding, yet remains competitive in shorter, visual tasks \cite{ataallah2024infinibench}. This may also help explain why simpler prompting strategies (e.g., zero-shot) sometimes outperform few-shot approaches for models like GPT-4o and Gemini, as additional examples can introduce noise or reduce alignment with the task objective.
\subsection{Costing and Real-World Applicability}

We analyze the computational cost of our pipeline, including video description generation, thumbnail prompt construction, and classification. The dataset’s average video length was 11.73 minutes; videos exceeding 30 minutes were truncated to 29:55, resulting in a final mean of 10.21 minutes. Reported costs reflect all inputs and outputs, including subtitles, generated descriptions, and classification prompts.

\begin{table}[t]
\centering
\small
\setlength{\tabcolsep}{4pt} 
\begin{tabular}{p{3.2cm}p{2.2cm}c}
\hline
\textbf{Component} & \textbf{Model} & \textbf{Cost/Video (\$)} \\
\hline
Video Desc. + Classif. & Claude & 0.0419 \\
Video Desc. + Classif. & Gemini & 0.0161 \\
Video Description & Twelve Labs & 0.4370 \\
Thumbnail Classif. & GPT-4o-mini & 0.0070 \\
Thumbnail Classif. & GPT-4o & 0.0529 \\
\hline
\end{tabular}
\caption{Cost comparison across pipeline components.}
\label{tab:costs}
\end{table}

Table~\ref{tab:costs} shows that Gemini provides the most cost-efficient end-to-end pipeline, while Claude achieves higher performance at a moderate increase in cost. Twelve Labs offers high-quality video understanding but is significantly more expensive, making it less practical for large-scale deployment without enterprise pricing~\cite{twelvelabs2024pricing}. Important to note, since Twelve Labs experiments were conducted using platform-provided credits, we report costs based on publicly available base pricing. Their enterprise-tier pricing will substantially reduce the rates at scale~\cite{twelvelabs2024pricing}. For classification, lightweight models such as GPT-4o-mini reduce costs by an order of magnitude compared to GPT-4o, with minimal degradation in performance, making them preferable for scalable systems.

To contextualize deployment costs, we estimate platform-scale expenses using YouTube’s reported upload volume of over 20 million videos per day (approximately 600 million per month) \cite{youtube_press}. At this scale, the full pipeline would cost approximately \$25.1M/month using Claude and \$9.7M/month using Gemini. Compared to YouTube’s estimated \$5B monthly revenue \cite{variety_youtube}, these costs represent approximately 0.50\% and 0.19\% of revenue, respectively, indicating that large-scale deployment remains economically feasible. Finally, LLM costs continue to decline. The subsequent GPT-4o release (\texttt{gpt-4o-2024-08-06}) halved input token rates and reduced output costs by 33\%. Techniques such as batch processing and prompt caching will further reduce operational costs and latency~\cite{openai2024batchguide,anthropic2024messagebatches,anthropic2024promptcaching,google2024contextcache,openai2024promptcaching,google2024batchpredictiongemini}, improving real-world feasibility.

\vspace{0.5em}
\noindent\textbf{Real-World Applicability.} Our pipeline integrates into YouTube’s infrastructure as a lightweight pre-upload layer. It processes the thumbnail, subtitles, and video description, and classifies thumbnails as misleading or not. If misleading, uploads can be blocked or flagged for review, enabling proactive moderation. Although built for YouTube, the modular design supports adaptation to other platforms such as TikTok, Instagram Reels, and Dailymotion. Future work will explore broader content categories and cross-platform generalization.

\subsection{Analysis of Open-Source Vision-Language Models}
To evaluate the general effectiveness of our framework, we additionally tested two open-source vision-language models, LLaVA-1.5 and Qwen-2.5-VL, under zero-shot, fixed few-shot, and dynamic few-shot prompting settings.

Overall, dynamic few-shot prompting consistently produced the strongest results for both models. LLaVA-1.5 improved from 33.5\% accuracy in the zero-shot setting to 67.3\% with dynamic retrieval-based prompting, while Qwen2.5-VL-7B-Instruct improved from 64.5\% to 74.0\%. These gains suggest that providing semantically relevant examples at inference time substantially improves the ability of open-source models to reason about whether thumbnails accurately reflect video content.

LLaVA-1.5 struggled considerably in the zero-shot setting, frequently misclassifying non-misleading thumbnails as misleading and generating verbose or inconsistent explanations. While fixed few-shot examples improved calibration, dynamic few-shot prompting produced the largest improvement by reducing incoherent outputs and improving reasoning consistency. In contrast, Qwen2.5-VL-7B-Instruct demonstrated much stronger zero-shot performance and more reliable multimodal reasoning from the outset. For Qwen as well, dynamically retrieved examples outperformed static examples, suggesting that adaptive contextual retrieval is more effective than generic fixed demonstrations.

Despite these improvements, both open-source models continued to struggle with grounding thumbnails in the broader video context. The dominant source of error across nearly all settings came from contextual reasoning failures, where models identified thumbnail elements but failed to determine whether they accurately represented the video content. For LLaVA-1.5, these failures accounted for 43.5\% of predictions in the zero-shot setting and remained high even with dynamic few-shot prompting (27.4\%). Similarly, Qwen2.5-VL-7B-Instruct exhibited contextual reasoning failure rates of 27.8\% in zero-shot and 24.5\% with dynamic few-shot prompting. 

LLaVA-1.5 also produced a noticeable number of incoherent outputs in earlier prompting settings, while fixed few-shot prompting slightly increased contextual reasoning errors for Qwen2.5-VL-7B-Instruct, suggesting that static examples can introduce biases for already capable models.

\vspace{0.5em}

\noindent \textbf{Open-Source vs. Proprietary Models.} While retrieval-augmented prompting substantially improved the performance of open-source vision--language models, a considerable gap remained between open-source and proprietary systems. The strongest open-source configuration, Qwen2.5-VL-7B-Instruct with dynamic few-shot prompting, achieved 74.0\% accuracy. All proprietary models outperformed the open-source models, with Claude 3.5 Sonnet achieving 93.8\% accuracy, GPT-4o-mini 84.8\%, Gemini 1.5 Flash 82.8\%, and GPT-4o 78.6\%. Notably, even the lowest-performing proprietary model here (GPT-4o 78.6\%) exceeded the best-performing open-source configuration by 4.6\%, while Claude outperformed Qwen2.5-VL-7B-Instruct by nearly 20\%.

These differences suggest that proprietary multimodal systems currently retain stronger capabilities for contextual reasoning and semantic alignment between thumbnails and video content. Open-source models frequently identified visual elements correctly but struggled to determine whether those elements accurately reflected the underlying video narrative, particularly in cases involving exaggerated thumbnails, celebrity manipulation, or culturally contextual clickbait. Nevertheless, the substantial gains achieved through dynamic few-shot prompting indicate that open-source models remain promising for accessible and reproducible moderation pipelines, especially as retrieval-augmented and multimodal reasoning capabilities continue to improve.
\subsection{Human Baseline} 
To contextualize model performance, we conducted a small-scale human baseline evaluation under the same multimodal input conditions used for the LLMs (thumbnail, subtitles, and video description). Five annotators each labeled 50 videos sampled from the dataset, requiring approximately 30 minutes on average. Across all annotations, human evaluators achieved 71.6\% accuracy and an F1 score of 0.670. These results suggest that identifying misleading thumbnails from limited multimodal context remains a challenging task even for human evaluators, while closed source LLMs substantially exceeded this baseline.

%% file: checker.tex
A central goal of our evaluation was to assess whether a modern LLM used in a zero-training, inference-only setting could outperform specialized multimodal pipelines designed for misleading thumbnail detection. To this end, we compared our best-performing setup, Claude 3.5 Sonnet with dynamic few-shot prompting, against CHECKER \cite{xie2021checker}. We selected CHECKER because it is the only prior model that directly addresses misleading thumbnails; other clickbait work focuses on text/metadata. We used CHECKER as released, following the authors’ default training setup. CHECKER fuses thumbnail and title features using advanced pooling mechanisms (e.g., Block, Mutan, MFH) and incorporates co-teaching to mitigate label noise. Its strongest configuration (Block pooling, $\tau = 0.30$) achieved an F1 score of 0.7153 on its publicly made available 197-video test set. However, when evaluated without weak supervision signals (i.e., generated labels), performance declined to 0.6538.
In contrast, Claude required no fine-tuning or supervision and achieved an F1 score of 0.7227 on the same CHECKER test set, outperforming CHECKER’s best result. Claude also surpassed several vision-language transformer baselines, including VisualBERT (0.6722), LXMERT (0.6640), and UNITER (0.6554), which were pretrained for multimodal alignment but struggled with the abstract or stylized nature of YouTube thumbnails. A traditional logistic regression baseline, using concatenated visual and textual embeddings, performed worst (0.4912 without, 0.5986 with generated labels), reflecting the limitations of shallow, non-interactive architectures.
These results demonstrate that prompt-driven LLMs can match or exceed the performance of supervised, domain-specific models, offering a flexible and training-free alternative for content moderation tasks, especially when guided by structured reasoning and contextual inputs.


%% file: ablation.tex
To evaluate the individual contributions of different textual modalities in our LLM-based detection pipeline, we conducted an ablation study using Claude 3.5 Sonnet in a zero-shot setting. The goal was to isolate the impact of subtitles and video descriptions on classification performance, while maintaining a consistent prompt structure. We performed the ablation only in the zero-shot setting to avoid altering few-shot exemplars, which rely on both subtitles and descriptions. Modifying these would introduce confounding factors, undermining the validity of the comparison.

We evaluated four input variants. The \textbf{Claude zero-shot} configuration uses the full input, consisting of the thumbnail image, video description, and subtitle transcript. \textbf{ABL-NS} removes subtitle information and relies only on the thumbnail and description. \textbf{ABL-ND} excludes the description and instead combines the thumbnail with subtitle text. Finally, \textbf{ABL-NDS} represents the most restricted setting and uses the thumbnail image alone, without access to either the description or subtitles.

\begin{table}[h]
\centering
\resizebox{\columnwidth}{!}{
\begin{tabular}{ccccc}
\hline
\textbf{Metric} & \textbf{ABL-NDS} & \textbf{ABL-ND} & \textbf{ABL-NS} & \textbf{Claude} \\
& & & & \textbf{Zero Shot} \\
\hline
Accuracy    & 0.8780 & 0.9077 & 0.9076 & 0.8920 \\
Recall      & 0.8010 & 0.8987 & 0.8856 & 0.8430 \\
Precision   & 0.9348 & 0.9016 & 0.9079 & 0.9240 \\
Specificity & 0.9487 & 0.9155 & 0.9258 & 0.9360 \\
\hline
\end{tabular}
}
\caption{Ablation Results – Claude 3.5 (Zero-Shot Setting)}
\label{tab:ablation_results}
\end{table}

As shown in Table~\ref{tab:ablation_results}, \textbf{Claude-Zero Shot} (the full-input configuration) yields the most balanced performance. Notably, \textbf{ABL-NS} (no subtitles) achieves similar accuracy and recall, indicating that descriptions alone often provide sufficient structured context for effective reasoning. Removing both textual modalities, \textbf{ABL-NDS}, leads to the weakest performance. Although specificity is highest, this likely reflects a conservative bias from lacking contextual input. In several MTV cases, the model refused classification entirely, citing ethical concerns (e.g., “I do not feel comfortable analyzing this type of sensationalized content...”). These were excluded from metric calculations. In \textbf{ABL-ND} (no description), the model relies on subtitles and performs well overall. However, missing or low-quality subtitles occasionally resulted in classification refusals, also excluded from reported metrics.


Overall, the results show that subtitles and descriptions offer complementary benefits. Subtitles improve the detection of specific misleading claims, while descriptions provide thematic grounding. Depending on the application, one may prioritize the higher recall of ABL-NS or the precision and specificity of the full-input Claude Zero-Shot configuration.\\\\

%% file: limitations.tex
\noindent
While our results demonstrate strong performance in detecting misleading thumbnail videos, the limitations of our system should be considered.

\vspace{0.5em}
\noindent
\textbf{Model Alignment and Description Generation.}
Our pipeline relies on Claude-generated video descriptions for open source models and uses Claude for downstream classification. Although this does not introduce direct leakage, it may create a self-reinforcement effect, where the model is advantaged when reasoning over representations it produced. This design ensures consistent, high-quality inputs but may bias results in favor of Claude-based pipelines. Due to resource constraints, we were unable to run all preprocessing pipelines for each model; a full cross-model ablation remains future work.

\vspace{0.5em}
\noindent
\textbf{Safety Filtering Effects.}
Different model APIs enforce varying safety filters, which can result in certain videos being excluded. Consequently, models may be evaluated on slightly different subsets, affecting direct comparability and observed failure modes.

\vspace{0.5em}
\noindent
{\textbf{Resolution and Data Constraints.}
All videos were processed at 360p and truncated to 30 minutes for scalability. While we did not observe substantial qualitative differences on a small higher-resolution subset, these choices may omit fine-grained signals and introduce bias in the dataset.

\vspace{0.5em}
\noindent
\textbf{Frame Sampling Strategy.}
We use 20 evenly spaced frames per video, which provides broad coverage but may miss short, temporally localized contradictions. More adaptive strategies, such as keyframe-based sampling, may better capture such cases.

%% file: discussion.tex
\label{sec:discussion}
The deployment of such a system by video platform providers could significantly enhance content moderation efforts. However, both challenges and opportunities would need to be carefully considered.

\vspace{0.5em}
\noindent
\textbf{False Positive Mitigation.} While our LLM-based approach demonstrates high precision, even a small fraction of false positives could impact legitimate content creators. To address this, platforms could implement a multi-stage review process where flagged thumbnails undergo human review before any action is taken.

\vspace{0.5em}
\noindent
\textbf{Transparency and Appeals.} Clear communication about the use of AI-assisted moderation and an efficient appeals process would be crucial to maintain user trust and provide recourse for incorrectly flagged content.

\vspace{0.5em}
\noindent
\textbf{Cultural and Linguistic Context.} As Mohan and Punathambekar \cite{mohan2019localizing} highlight YouTube’s struggle to balance global and local strategies in linguistically diverse regions, LLMs may face similar challenges in regions lacking sufficient linguistic or cultural data, potentially impacting the accuracy of thumbnail classification.

\vspace{0.5em}
\noindent
\textbf{Adaptive Systems.} Given the evolving nature of online content, integrating this system into a continuous learning pipeline enables ongoing refinement via updated LLMs, in-context learning, or fine-tuning to address emerging misleading content.

\vspace{0.5em}
\noindent
\textbf{Regulatory Compliance.} 
As regulations like the EU’s Digital Services Act (DSA) demand greater transparency and accountability \cite{eu_digital_services_act}, LLM-assisted moderation can help platforms like YouTube detect and remove harmful content, while enabling regulators to audit compliance and enforce policy standards.

%% file: related-work.tex
Prior work on detecting misleading content on platforms like YouTube has largely focused on videos and associated metadata such as tags and titles. UCNet \cite{palod2019misleading}, OVCP \cite{shang2019towards}, and Bajaj et al. \cite{bajaj2016disinformation} rely heavily on user engagement or metadata signals, limiting their use to post-hoc detection. These approaches do not address the visual-semantic alignment of thumbnails with content, a key focus of our work.

CHECKER \cite{xie2021checker} and BaitRadar \cite{gamage2021baitradar} move toward thumbnail-based analysis but either rely on weak heuristics or omit actual video content. Our comparison with CHECKER demonstrates that LLM-based pipelines outperform such approaches. Furthermore, limitations in dataset availability (as in the case of BaitRadar) and data quality (as in CHECKER, which relies on crowdsourced annotations) further constrain meaningful comparability.

Recent studies have explored LLMs for automated content analysis \cite{gilardi2023chatgpt, gonzalez2024benchmarking}, while moderation tools like PIXELMOD \cite{paudel2024pixelmod} emphasize visual content. These align with our use of LLMs and highlight a growing shift toward more semantic, context-aware moderation approaches.

\vspace{0.5em}
\noindent
\textbf{Our Contribution.}  
We present a large-scale, cross-country dataset with balanced annotations by trained evaluators and propose an LLM-based pipeline to assess semantic alignment between thumbnails and video content, addressing prior limitations in dataset design and detection methods.

%% file: conclusion.tex
This paper presented a comprehensive analysis of misleading video thumbnails on YouTube, leveraging a large dataset and advanced LLMs to improve existing detection methods. Our approach demonstrated higher accuracy compared to traditional techniques relying on metadata and user comments. Proprietary multimodal LLMs also consistently outperformed open-weight vision--language models, highlighting the current performance advantages of closed-source systems for complex multimodal reasoning tasks. The findings highlight the need for more robust and scalable solutions to mitigate misleading content on video platforms. We recommend that platforms like YouTube enhance their enforcement mechanisms and transparency to protect viewers from misleading thumbnails and improve content consumption experiences.

%% file: checklist.tex



\section{Paper Checklist}

\begin{enumerate}

\item For most authors...
\begin{enumerate}
    \item  Would answering this research question advance science without violating social contracts, such as violating privacy norms, perpetuating unfair profiling, exacerbating the socio-economic divide, or implying disrespect to societies or cultures?
    \answerTODO{Yes. The work advances understanding of visual misinformation and platform behavior using publicly available content, without collecting private user data or targeting specific individuals or groups.}
  \item Do your main claims in the abstract and introduction accurately reflect the paper's contributions and scope?
    \answerTODO{Yes. They accurately describe the scope, methodology, and empirical findings proving the claims accurately with results.}
   \item Do you clarify how the proposed methodological approach is appropriate for the claims made? 
    \answerTODO{Yes. We explain how large-scale measurement, multimodal analysis, and automated content analysis support the empirical claims. Moreover, due to increasing number of videos being uploaded, such an approach is necessary to automate misinformation detection efficiently.}
   \item Do you clarify what are possible artifacts in the data used, given population-specific distributions?
    \answerTODO{Yes. We discuss potential biases related to content popularity in various countries and regions and recommendation dynamics.}
  \item Did you describe the limitations of your work?
    \answerTODO{Yes, we discuss the limitations of our work while analyzing the results.}
  \item Did you discuss any potential negative societal impacts of your work?
    \answerTODO{Yes. We discuss potential negative impacts such as non misleading thumbnails in some cases getting marked as misleading.}
      \item Did you discuss any potential misuse of your work?
    \answerTODO{Yes, We talk about false positives but in particular there are no potential misuses.}
    \item Did you describe steps taken to prevent or mitigate potential negative outcomes of the research, such as data and model documentation, data anonymization, responsible release, access control, and the reproducibility of findings?
    \answerTODO{The proposed approach is to counter existing negative practices, hence there are no potential negative outcomes of our research. We have released data anonymously alongwith scripts for reproducibility of findings.}
  \item Have you read the ethics review guidelines and ensured that your paper conforms to them?
    \answerTODO{Yes}
\end{enumerate}

\item Additionally, if your study involves hypotheses testing...
\begin{enumerate}
  \item Did you clearly state the assumptions underlying all theoretical results?
    \answerTODO{Not applicable.}
  \item Have you provided justifications for all theoretical results?
    \answerTODO{Not applicable.}
  \item Did you discuss competing hypotheses or theories that might challenge or complement your theoretical results?
    \answerTODO{Not applicable.}
  \item Have you considered alternative mechanisms or explanations that might account for the same outcomes observed in your study?
    \answerTODO{Yes. We conducted ablation studies to observe different outcomes and effectiveness of the same method but with different inputs.}
  \item Did you address potential biases or limitations in your theoretical framework?
    \answerTODO{Not applicable.}
  \item Have you related your theoretical results to the existing literature in social science?
    \answerTODO{Not applicable.}
  \item Did you discuss the implications of your theoretical results for policy, practice, or further research in the social science domain?
    \answerTODO{Not applicable.}
\end{enumerate}

\item Additionally, if you are including theoretical proofs...
\begin{enumerate}
  \item Did you state the full set of assumptions of all theoretical results?
    \answerTODO{Not Applicable.}
	\item Did you include complete proofs of all theoretical results?
    \answerTODO{Not Applicable.}
\end{enumerate}

\item Additionally, if you ran machine learning experiments...
\begin{enumerate}
  \item Did you include the code, data, and instructions needed to reproduce the main experimental results (either in the supplemental material or as a URL)?
    \answerTODO{Yes. Code, datasets, and instructions are provided via an anonymized repository.}
  \item Did you specify all the training details (e.g., data splits, hyperparameters, how they were chosen)?
    \answerTODO{Not applicable. Used LLMs in default settings without training.}
     \item Did you report error bars (e.g., with respect to the random seed after running experiments multiple times)?
    \answerTODO{We ran on a smaller subset multiple times due to limited resources and got consistent results.}
	\item Did you include the total amount of compute and the type of resources used (e.g., type of GPUs, internal cluster, or cloud provider)?
    \answerTODO{Not applicable but we did a cost analysis.}
     \item Do you justify how the proposed evaluation is sufficient and appropriate to the claims made? 
    \answerTODO{Yes. The evaluation aligns with the measurement and analysis goals of the paper.}
     \item Do you discuss what is ``the cost`` of misclassification and fault (in)tolerance?
    \answerTODO{Yes. We have a paragraph on such implications.}
  
\end{enumerate}

\item Additionally, if you are using existing assets (e.g., code, data, models) or curating/releasing new assets, \textbf{without compromising anonymity}...
\begin{enumerate}
  \item If your work uses existing assets, did you cite the creators?
    \answerTODO{No external datasets were used.}
  \item Did you mention the license of the assets?
    \answerTODO{Not applicable.}
  \item Did you include any new assets in the supplemental material or as a URL?
    \answerTODO{Only URL included is of our repository containing dataset and scripts.}
  \item Did you discuss whether and how consent was obtained from people whose data you're using/curating?
    \answerTODO{Not applicable.}
  \item Did you discuss whether the data you are using/curating contains personally identifiable information or offensive content?
    \answerTODO{Not applicable.}
\item If you are curating or releasing new datasets, did you discuss how you intend to make your datasets FAIR?
\answerTODO{Yes, the dataset has been made public, and is FAIR.}
\item If you are curating or releasing new datasets, did you create a Datasheet for the Dataset? 
\answerTODO{We created our own dataset, and made it public.}
\end{enumerate}

\item Additionally, if you used crowdsourcing or conducted research with human subjects, \textbf{without compromising anonymity}...
\begin{enumerate}
  \item Did you include the full text of instructions given to participants and screenshots?
    \answerTODO{Yes, we had 2 annotators which were provided a codebook and sample thumbnails which are uploaded on our anonymized GitHub repo. The 5 survey takers were also given entire list of instructions in the survey.}
  \item Did you describe any potential participant risks, with mentions of Institutional Review Board (IRB) approvals?
    \answerTODO{Not applicable.}
  \item Did you include the estimated hourly wage paid to participants and the total amount spent on participant compensation?
    \answerTODO{Not applicable.}
   \item Did you discuss how data is stored, shared, and deidentified?
   \answerTODO{Not applicable.}
\end{enumerate}

\end{enumerate}

%% file: appendix.tex
\appendix
\section{Appendix}

\subsection{Zero-shot Prompt}
\begin{quote}
\textbf{Task:} Analyze the provided information about a YouTube video and determine whether its thumbnail is misleading or not misleading. You will be given the following information:
\begin{itemize}
\item The video's thumbnail
\item The video's subtitles
\item A text description of the video content
\end{itemize}

\textbf{Steps to Follow:}
\begin{enumerate}
\item Carefully examine the thumbnail.
\item Read through the video subtitles and content description.
\item Compare the thumbnail to the actual video content.
\item Determine if the thumbnail accurately represents the video's main topic or content.
\item Assess whether the thumbnail uses exaggeration, false promises, or clickbait tactics.
\item Categorize the thumbnail as either \textit{``Misleading"} or \textit{``Not Misleading"}.
\item Provide a brief explanation for your decision.
\end{enumerate}
Analyze the provided information and categorize the thumbnail as ``Misleading" or ``Not Misleading", followed by a brief explanation for your decision.

\textbf{Inputs:} \\
Video Description: \{\texttt{video\_description}\} \\
Video Subtitles: \{\texttt{video\_subtitles}\}
\end{quote}

\subsection{Fixed Few-shot Examples}
\begin{quote}
\textbf{Example 1:} \\
\textbf{Thumbnail:} A person holding a stack of \$100 bills with the text “I made \$10,000 in one day!” \\
\textbf{Subtitles:} “In this video, I’ll share my experience of how I earned \$500 in a week through freelancing.” \\
\textbf{Video Description:} The creator discusses freelancing opportunities and shares how they earned \$500 in their first week. \\
\textbf{Categorization:} Misleading \\
\textbf{Explanation:} The thumbnail exaggerates the earnings (i.e., \$10,000 in a day) compared to the actual content (\$500 in a week), using clickbait tactics.

\vspace{1mm}

\textbf{Example 2:} \\
\textbf{Thumbnail:} A smiling chef holding a plate of pasta with the text “Easy 15-minute pasta recipe.” \\
\textbf{Subtitles:} “Today, we’re making a quick and delicious pasta dish that takes only 15 minutes to prepare.” \\
\textbf{Video Description:} The video demonstrates a step-by-step pasta recipe with a 15-minute prep time. \\
\textbf{Categorization:} Not Misleading \\
\textbf{Explanation:} The thumbnail accurately represents the content and preparation time.

\end{quote}

\subsection{Example Claude Output} \label{appendix:claude_response}
Below is an example of Claude’s output when evaluating a thumbnail featuring a “banana cobra” and a “Spider-Man snail”:
\begin{quote}
Based on the provided information, I would categorize this thumbnail as Misleading. Here's my explanation:\\
\textbf{Thumbnail content:} The thumbnail shows two unrelated images—a cobra snake made from a banana, and a snail-like creature with Spider-Man's appearance.\\
\textbf{Video content:} The video description and subtitles indicate that this is a compilation of various interesting and unusual scenes, including sunglasses, exercise equipment mishaps, recycling facilities, people walking on hot coals, and many other diverse topics. None of these match the images in the thumbnail.\\
\textbf{Lack of representation:} The thumbnail images are not mentioned anywhere in the video description or subtitles. There's no indication that the video actually includes content related to banana snakes or Spider-Man snails.\\
\textbf{Misleading expectations:} Viewers expecting to see content related to the thumbnail images would likely be disappointed, as the video seems to cover a wide range of unrelated topics.
The thumbnail is clearly designed to grab attention with its unusual and striking images, but it does not accurately represent the diverse compilation of scenes and activities actually present in the video. This discrepancy between the thumbnail and the actual content makes it misleading to potential viewers.
\end{quote}





\subsection{Illustrative Annotation Examples}
\begin{figure}[H]
    \centering
    \includegraphics[width=\linewidth]{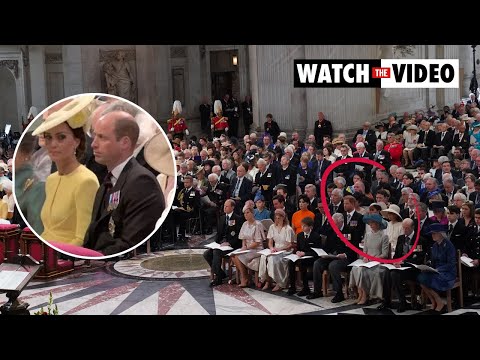}
    \caption{Borderline non-misleading example where visual exaggeration and selective emphasis are used without substantially misrepresenting the video content.}
    \label{fig:nmtv_thumbnail_examples}
\end{figure}

\begin{figure}[H]
    \centering
    \includegraphics[width=\linewidth]{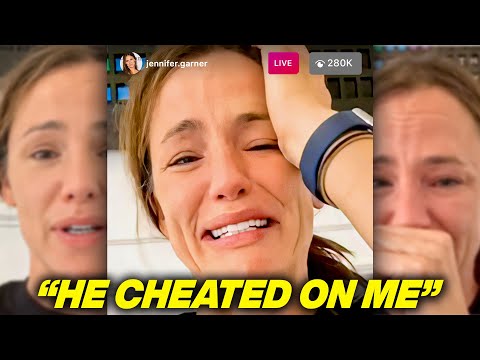}
    \caption{Misleading thumbnail containing fabricated implications and emotionally manipulative framing not supported by the video content.}
    \label{fig:mtv_thumbnail_examples}
\end{figure}

Figures~\ref{fig:nmtv_thumbnail_examples} and~\ref{fig:mtv_thumbnail_examples} present illustrative borderline examples used during annotation. Figure~\ref{fig:nmtv_thumbnail_examples} shows a non-misleading thumbnail where the creator visually exaggerates and highlights reactions from a real event to increase viewer attention. Although the thumbnail is dramatized through zooming and selective emphasis, the depicted event genuinely appears in the video and does not substantially misrepresent the underlying content.

In contrast, Figure~\ref{fig:mtv_thumbnail_examples} shows a misleading thumbnail falsely implying that actress Jennifer Garner publicly stated “He cheated on me” during an emotional Instagram Live session. While the thumbnail uses emotionally charged imagery and fabricated text to suggest a dramatic revelation, the video itself does not contain such claims. This example illustrates the distinction used during annotation between acceptable exaggeration and genuine thematic mismatch or fabricated implication.

\subsection{Supplementary Results and Data}
\begin{table}[ht]
\centering
\begin{tabular}{|c|c|}
\hline
\textbf{Model}      & \textbf{Videos Processed} \\ \hline
Claude              & 2759                      \\ \hline
Gemini              & 2135                      \\ \hline
GPT-4o-mini \& Twelve Labs & 2769              \\ \hline
GPT-4o  \& Twelve Labs     & 2749              \\ \hline
\end{tabular}
\caption{Number of Videos Processed by Each Model}
\label{tab:videos_processed}
\end{table}





\begin{figure}[h] 
\centering 
\includegraphics[width=\linewidth]{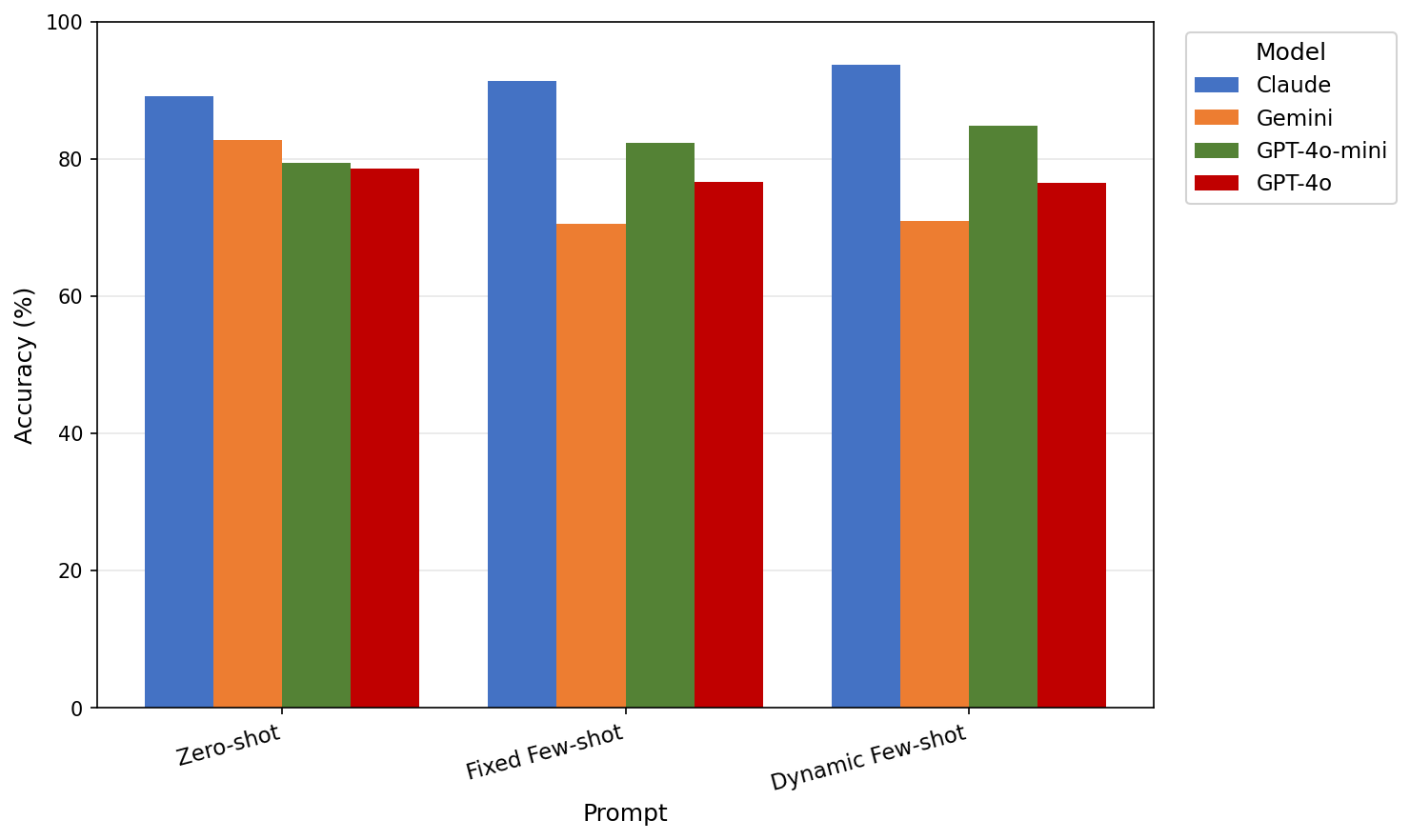} 
\caption{Graph comparing model performances across prompts} 
\label{fig:overall_accuracy} 
\end{figure}


\begin{figure}[h] 
\centering 
\includegraphics[width=\linewidth]{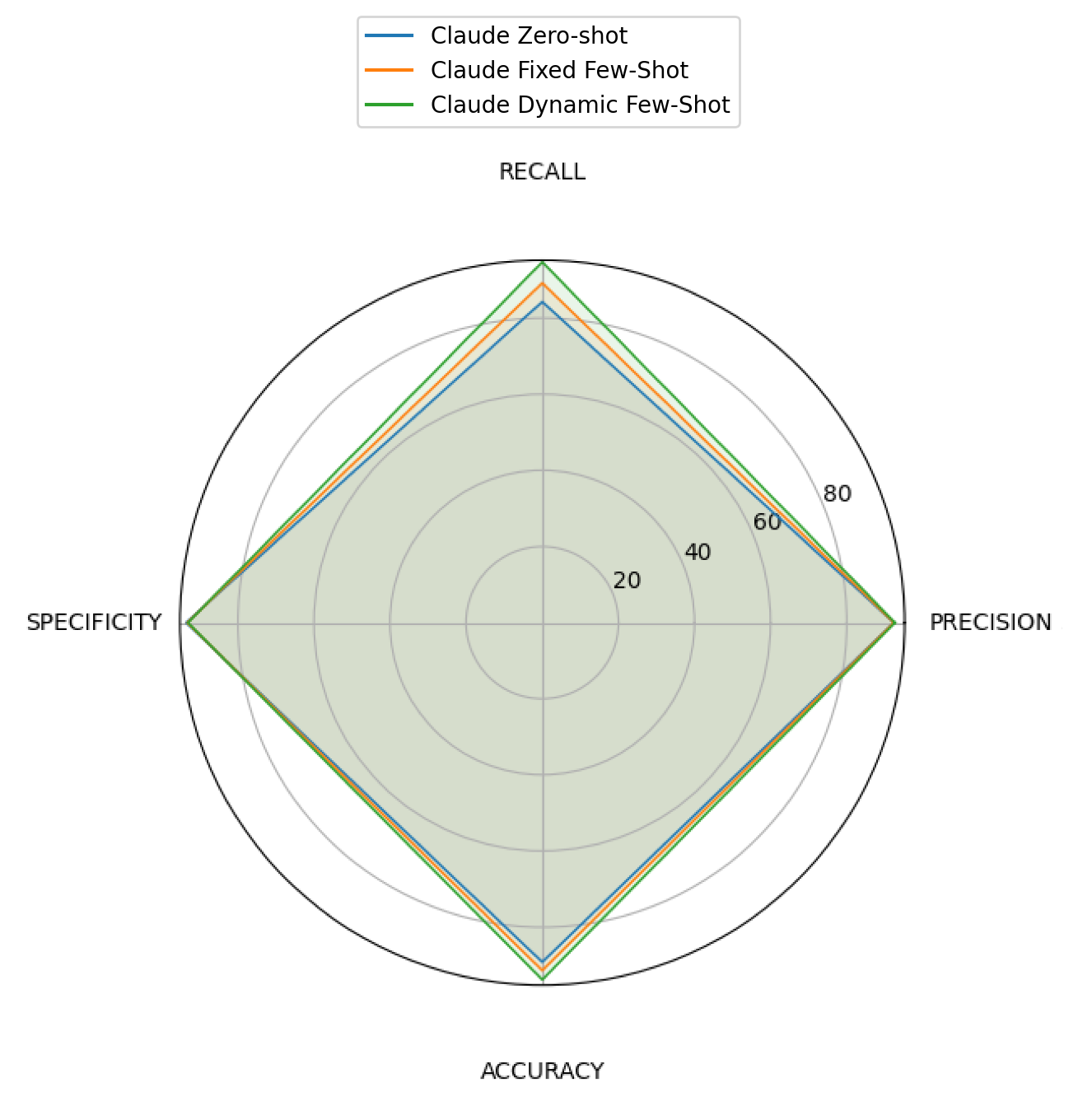} 
\caption{Radar plot for Claude for all prompts} 
\label{fig:radar_claude} 
\end{figure}


\begin{figure}[h] 
\centering 
\includegraphics[width=\linewidth]{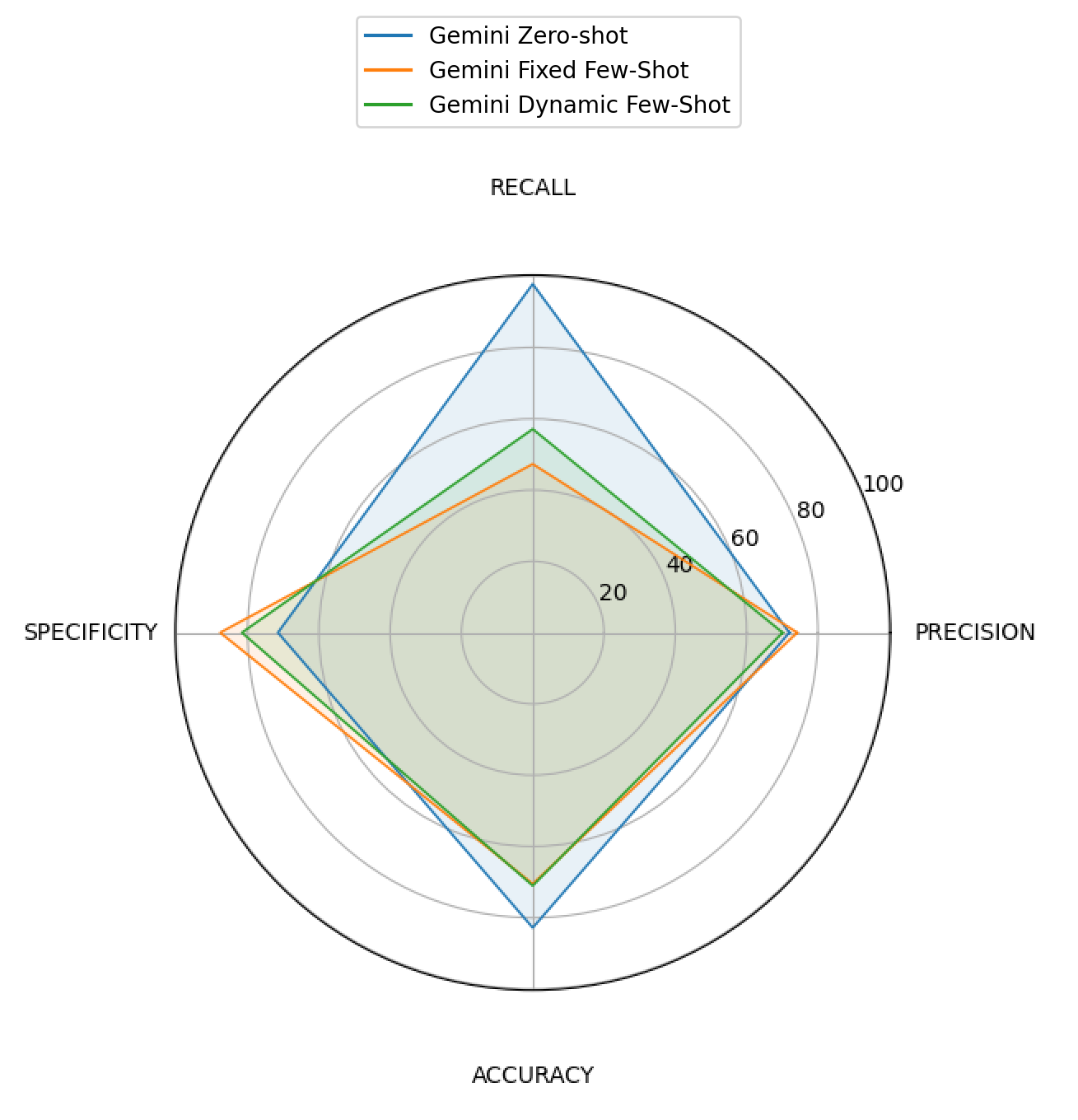} 
\caption{Radar plot for Gemini for all prompts} 
\label{fig:radar_gemini} 
\end{figure}


\begin{figure}[h] 
\centering 
\includegraphics[width=\linewidth]{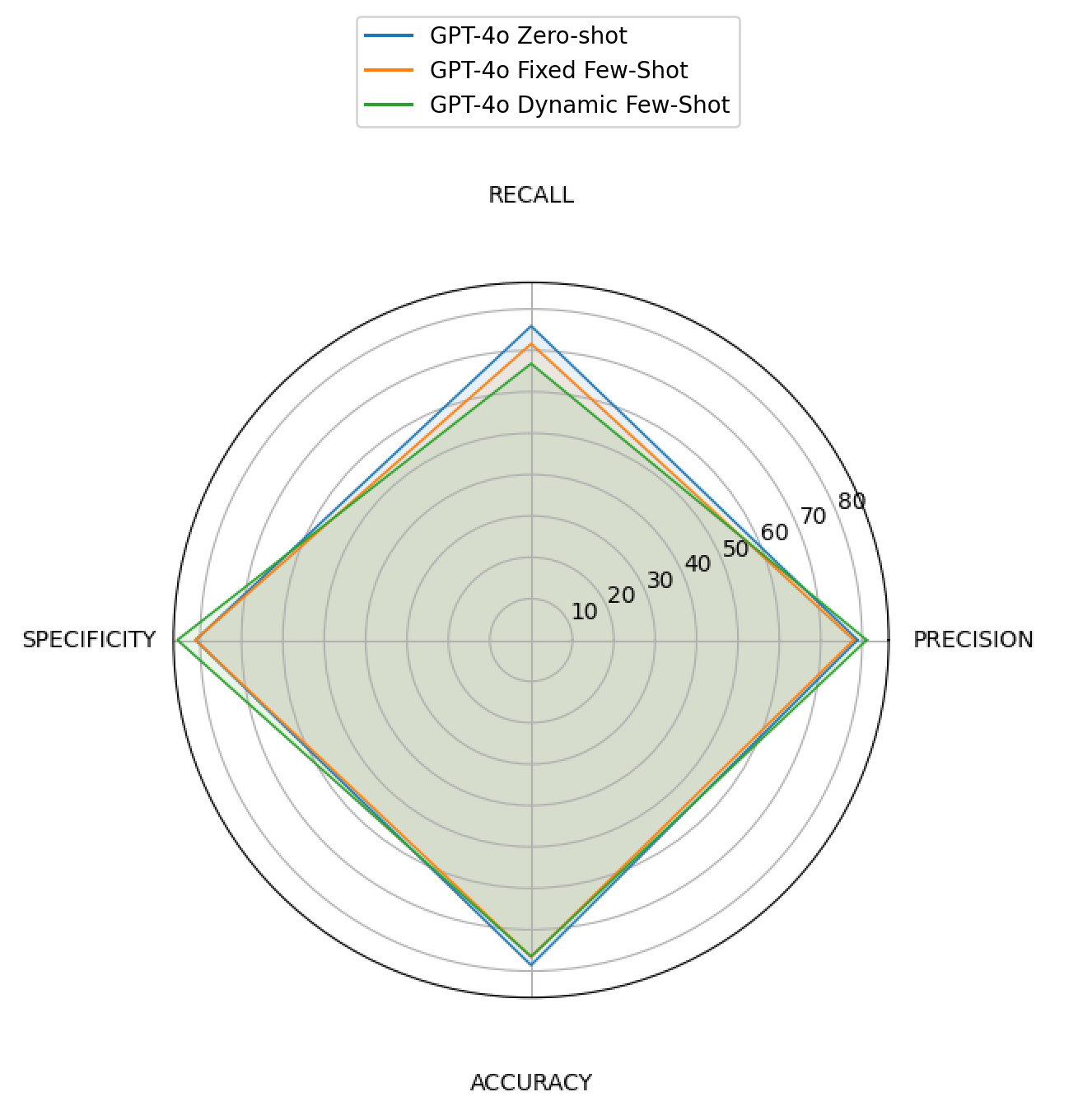} 
\caption{Radar plot for GPT-4o for all prompts} 
\label{fig:radar_gpt4o} 
\end{figure}

\begin{figure}[h] 
\centering 
\includegraphics[width=\linewidth]{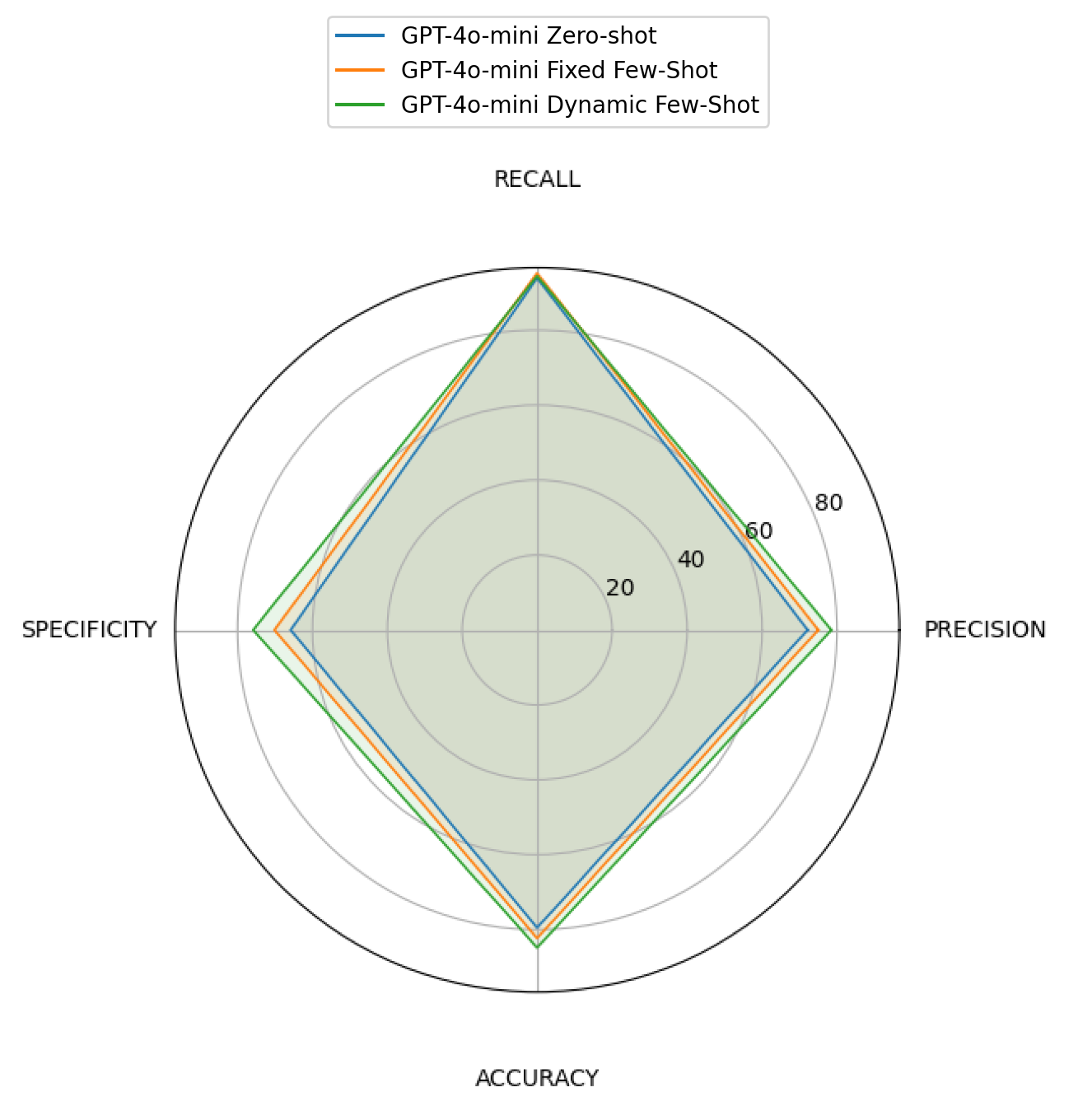} 
\caption{Radar plot for GPT-4o-mini for all prompts} 
\label{fig:radar_gpt4omini} 
\end{figure}

\begin{figure}[h] 
\centering 
\includegraphics[width=\linewidth]{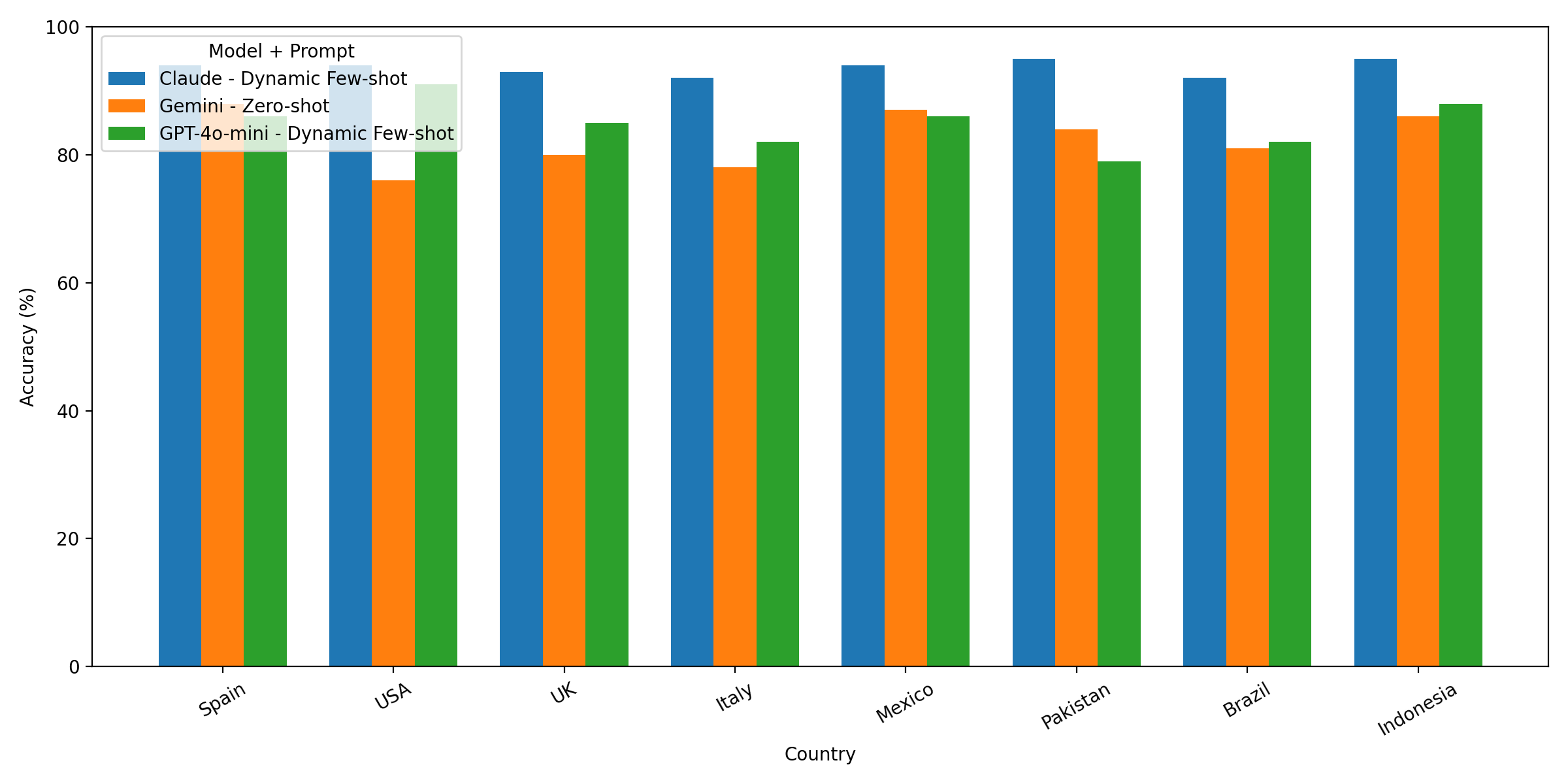} 
\caption{Top 3 Models with their Highest Individual Accuracy} 
\label{fig:bestaccuracy_top3models} 
\end{figure}

\begin{figure}[h] \centering \includegraphics[width=\linewidth]{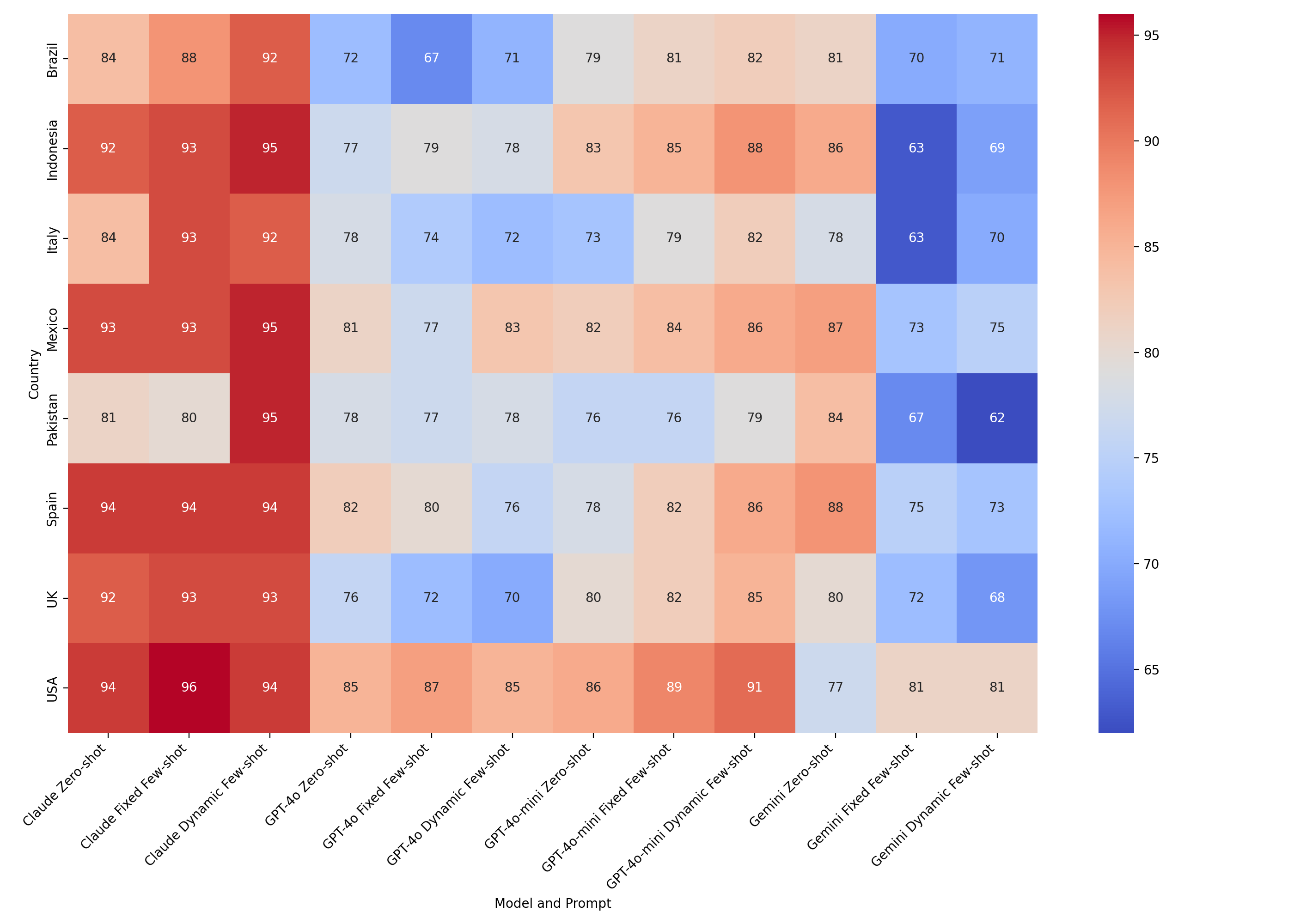} \caption{Heatmap for Model Accuracies Across Prompts for Each Country} \label{fig
} \end{figure}

\begin{figure}[h]
    \centering
    \includegraphics[width=\linewidth]{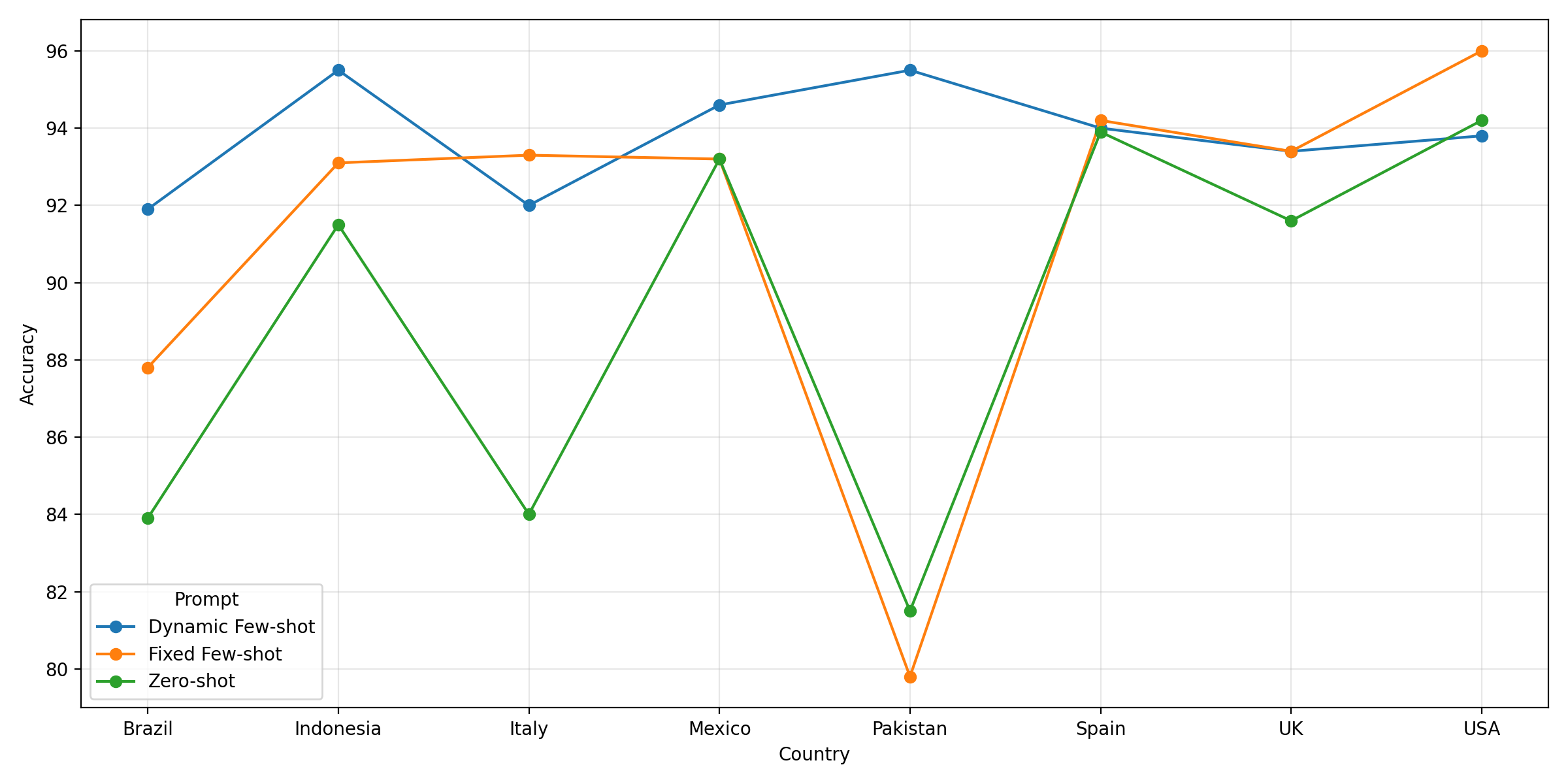}
    \caption{Claude Accuracies Across Prompts for Each Country}
    \label{fig1}
\end{figure}